\documentclass[sigconf]{acmart}
\hypersetup{draft}
\usepackage{tcolorbox}
\usepackage{threeparttable}

\usepackage{ifthen}
\newboolean{showcomments}
\setboolean{showcomments}{false}

\newcommand{\TODO}[1]{\noindent{\textbf{TODO: #1}}}
\ifthenelse{\boolean{showcomments}}
  {\newcommand{\nb}[2]{
    \fbox{\bfseries\sffamily\scriptsize\color{red}#1}
    {\sf\small$\blacktriangleright${\color{red}\textit{#2}}$\blacktriangleleft$}
   }
   \newcommand\TBDSEC[2]{\TODO{TBD} \textcolor{#2}{\lipsum[1-#1]}}
  }
  {\newcommand{\nb}[2]{}
   
   \newcommand\TBDSEC[2]{}
  }

\AtBeginDocument{%
  \providecommand\BibTeX{{%
    \normalfont B\kern-0.5em{\scshape i\kern-0.25em b}\kern-0.8em\TeX}}}

\copyrightyear{2020}
\acmYear{2020}
\setcopyright{acmcopyright}
\acmConference[MSR '20]{17th International Conference on Mining Software Repositories}{October 5--6, 2020}{Seoul, Republic of Korea}
\acmBooktitle{17th International Conference on Mining Software Repositories (MSR '20), October 5--6, 2020, Seoul, Republic of Korea}
\acmPrice{15.00}
\acmDOI{10.1145/3379597.3387467}
\acmISBN{978-1-4503-7517-7/20/05}

\begin{document}

%%
%% The "title" command has an optional parameter,
%% allowing the author to define a "short title" to be used in page headers.
\title{On the Prevalence, Impact, and Evolution of SQL Code Smells in Data-Intensive Systems}

%%
%% The "author" command and its associated commands are used to define
%% the authors and their affiliations.
%% Of note is the shared affiliation of the first two authors, and the
%% "authornote" and "authornotemark" commands
%% used to denote shared contribution to the research.

\author{Biruk Asmare, Muse}
\affiliation{
  \institution{Polytechnique Montréal}
  %\streetaddress{1 Th{\o}rv{\"a}ld Circle}
  %\city{Hekla}
  %\country{Iceland}
  }
\email{biruk-asmare.muse@polymtl.ca}
\author{Mohammad Masudur Rahman}
\affiliation{
  \institution{Polytechnique Montréal}
  %\streetaddress{1 Th{\o}rv{\"a}ld Circle}
  %\city{Hekla}
  %\country{Iceland}
  }
\email{masud.rahman@polymtl.ca}
\author{Csaba Nagy}
\affiliation{
  \institution{Software Institute\\Università della Svizzera italiana}
  %\streetaddress{1 Th{\o}rv{\"a}ld Circle}
  %\city{Hekla}
  %\country{Iceland}
  }
\email{csaba.nagy@usi.ch}
\author{Anthony Cleve}
\affiliation{
  \institution{Namur Digital Institute\\University of Namur}
  %\streetaddress{1 Th{\o}rv{\"a}ld Circle}
  %\city{Hekla}
  %\country{Iceland}
  }
\email{anthony.cleve@unamur.be}
\author{Foutse Khomh}
\affiliation{
  \institution{Polytechnique Montréal}
  %\streetaddress{1 Th{\o}rv{\"a}ld Circle}
  %\city{Hekla}
  %\country{Iceland}
  }
\email{foutse.khomh@polymtl.ca}
\author{Giuliano Antoniol}
\affiliation{
  \institution{Polytechnique Montréal}
  %\streetaddress{1 Th{\o}rv{\"a}ld Circle}
  %\city{Hekla}
  %\country{Iceland}
  }
\email{antoniol@ieee.org}
%%
%% By default, the full list of authors will be used in the page
%% headers. Often, this list is too long, and will overlap
%% other information printed in the page headers. This command allows
%% the author to define a more concise list
%% of authors' names for this purpose.
\renewcommand{\shortauthors}{}

%%
%% The abstract is a short summary of the work to be presented in the
%% article.
\begin{abstract}
Code smells indicate software design problems that harm software quality.
Data-intensive systems that frequently access databases often suffer from SQL code smells besides the traditional smells. While there have been extensive studies on traditional code smells, recently, there has been a growing interest in SQL code smells. In this paper, we conduct an empirical study to investigate the prevalence and evolution of SQL code smells in open-source, data-intensive systems. We collected 150 projects and examined both traditional and SQL code smells in these projects. Our investigation delivers several important findings. First, SQL code smells are indeed prevalent in data-intensive software systems. Second, SQL code smells have a weak co-occurrence with traditional code smells. Third, SQL code smells have a weaker association with bugs than that of traditional code smells. Fourth, SQL code smells are more likely to be introduced at the beginning of the project lifetime and likely to be left in the code without a fix, compared to traditional code smells. Overall, our results show that SQL code smells are indeed prevalent and persistent in the studied data-intensive software systems. Developers should be aware of these smells and consider detecting and refactoring SQL code smells and traditional code smells separately, using dedicated tools.
%, which implies that SQL code smells should be detected and refactored separately using dedicated tools. %more sophisticated tools. 
%Fourth, 88\% of SQL code smells were created in earliest releases of the system and persisted for a long time without getting fixed compared to 65\% for traditional code smells. \Foutse{this means that 65\% of traditional code smells were also created in the first version and persisted...is that correct? is that what you meant? yes} 

%the median survival time of SQL code smells (four years) is shorter than that of the traditional code smells (nine years) in our study, which is still a long time. 

%Developers should be aware of these smells and consider refactoring SQL code smells and traditional code smells separately. %them at the early stages to mitigate

%\Foutse{why performance and maintainability? did we assess connection with these quality attribute? we need to have established a clear link before making such claim!} %performance and maintainability issues.
 %Overall, .
\end{abstract}

%%
%% The code below is generated by the tool at http://dl.acm.org/ccs.cfm.
%% Please copy and paste the code instead of the example below.
%%
\begin{CCSXML}
<ccs2012>
   <concept>
       <concept_id>10011007.10011074.10011099.10011693</concept_id>
       <concept_desc>Software and its engineering~Empirical software validation</concept_desc>
       <concept_significance>300</concept_significance>
       </concept>
 </ccs2012>
\end{CCSXML}

\ccsdesc[300]{Software and its engineering~Empirical software validation}

%%
%% Keywords. The author(s) should pick words that accurately describe
%% the work being presented. Separate the keywords with commas.
\keywords{Code smells, database access, SQL code smells, data-intensive systems}

%%
%% This command processes the author and affiliation and title
%% information and builds the first part of the formatted document.
\maketitle

% -----------------------------------------------------------------------------------------------------------------------------------------------------------------

\section{Introduction}
In Software Engineering, \textit{smells} are \textit{poor} solutions to commonly occurring problems in a software system. They could be found within the design and implementation of the system, and are frequently referred to as design smells and code smells, respectively. Regardless of whether they originate from the design or the source code, previous work have shown that they can negatively affect the performance \cite{hecht2016empirical} or the maintainability \cite{khomh2009exploratory} of a software system. They should therefore be handled with special care, and refactored \cite{FowlerRefactoring} as soon as possible. %before they do more harm. 
We refer to these smells as \textit{traditional smells} throughout the paper. While there have been extensive studies on traditional smells \cite{khomh2009exploratory,shatnawi2006investigation,palomba2018diffuseness,johannes2019large}, recently, there has been a growing interest in a particular type of smells, namely \emph{SQL code smells} \cite{karwin2010sql,nagy2017static,de2019prevalence}.

SQL code smells are found within SQL code as a result of misuses in queries (e.g., \emph{Implicit Columns} \cite{nagy2017static}). Data-intensive software systems that frequently interact with databases are particularly prone to these smells. 
The SQL queries are often embedded in the application code and remain hidden from the developers, which makes it harder to spot mistakes in them.
As SQL is the principal language to communicate with relational databases, which represent the Top-5 databases according to the DB-Engine Ranking\footnote{\url{https://db-engines.com/en/ranking}}, many systems might suffer from such a type of smells.
Although traditional code smells are widely studied, there have been only a few studies on the prevalence and impact of SQL code smells \cite{de2019prevalence, Lyu2019, Sharma2018}. The objective of our study is to address this gap in the literature of code smells.

In this paper, we conduct an empirical study on SQL code smells, that investigates their (1) prevalence, (2) impact, (3) evolution and (4) co-occurrence with the traditional code smells within hundreds of open-source software systems. To the best of our knowledge, this is the first study investigating the prevalence, impact, and evolution of SQL code smells and also contrasting with the traditional 
%code 
smells. 
%First, it will help us to understand the co-occurrence and interaction of both types of smells and perform informed refactoring decisions. Second, it will give as a base line to study the impact and persistence of SQL code smells.

Our study relies on the analysis of 150 open-source software systems that manipulate their databases through popular database access APIs -- Android Database API, JDBC, JPA and Hibernate. We analysed the source code of each project and studied 19 traditional code smells using the DECOR tool \cite{gueheneuc2007ptidej} and 4 SQL code smells using the SQLInspect tool \cite{nagy2018sqlinspect}. We also collected bug-fixing and bug-inducing commits from each project using PyDriller \cite{PyDriller}. 

By analyzing the collected data, we answer the four following research questions:

\textbf{RQ1: What is the prevalence of SQL code smells across different application domains?}

We study the prevalence of SQL code smells in the selected software systems by categorizing them into four application domains -- Business, Library, Multimedia, and Utility. We find that SQL code smells are prevalent in all four domains. Some SQL code smells are more prevalent than others. 

\textbf{RQ2: Do traditional code smells and SQL code smells co-occur at class level?}

We investigate the co-occurrence of SQL code smells and traditional code smells using association rule mining. The results show that while some SQL code smells have statistically significant co-occurrence with traditional code smells, the degree of association is low.

\textbf{RQ3: Do SQL code smells co-occur with bugs?}

We investigate the potential impact of SQL code smells on software bugs by analysing their co-occurrences within the bug-inducing commits. We perform Cramer's V test of association and build a random forest model to study the impact of the smells on bugs. We find that there is a weak association between SQL code smells and software bugs. Some SQL code smells tend to show higher association with bugs compared to others.

\textbf{RQ4: How long do SQL code smells survive?}

We perform a survival analysis of SQL and traditional code smells using Kaplan-Meier survival curves to compare their survival time. It is interesting to know the lifespan of SQL code smells as software evolves. It indicates whether the smells stay longer without getting fixed or not. We find that the survival time of SQL code smells is higher compared to that of traditional code smells. Furthermore, significant portions of the SQL code smells are created at the very beginning and then persist in all subsequent versions of the systems. 

%\Foutse{it would be interesting to explain if the sql smells are being refactored or if they end up being remove just accidentally...for example by removing a chunk of code?}  

%\textbf{Structure of the paper:} In Section \ref{sec:background}, we provide background information about smells and about the statistical analysis techniques we used. We describe our research method in Section \ref{sec:method}. We discuss our findings in Section \ref{sec:find} and their implications in Section \ref{sec:implication}. The threats to validity of our study are identified in Section \ref{sec:threats}. In Section \ref{sec:related} we position our work with respect to the related literature. Section \ref{sec:conclusion} gives concluding remarks and outlines some future research directions.

% -----------------------------------------------------------------------------------------------------------------------------------------------------------------

\section{Background}
\label{sec:background}
\subsection{SQL Code Smells}\label{sec:sql-smells}
Besides many white papers and blog posts about common mistakes or bad practices in SQL queries \cite{RedGateSQLSmells}, an extensive catalogue of SQL code smells was published by Karwin in 2010 \cite{karwin2010sql}.%his book on SQL AntiPatterns 

There are also tools (e.g., TOAD and SQL Enlight) typically designed for database administrators that can statically analyse queries and identify common mistakes. These techniques require the SQL code as input. For our study, to investigate SQL code embedded in the source code of data-intensive systems, we rely on SQLInspect \cite{nagy2018sqlinspect}, a tool able to extract SQL code from Java applications and detect SQL code smells belonging to Karwin's catalogue. In the following, we briefly describe the SQL code smells that SQLInspect can detect.

\textbf{Implicit Columns} smell occurs when columns of a table are unnecessarily queried, e.g., the usage of \texttt{*} in the column list of a \texttt{SELECT} statement. Although it is fast to write, it may cause performance issues such as network bandwidth wastage or even more serious problems when the table column order is modified and the change is not propagated to the application code \cite{karwin2010sql}.

\textbf{Fear of the Unknown} is a smell that occurs due to improper handling of \texttt{NULL} values. \texttt{NULL} has a special meaning in relational databases as it indicates the absence of data, and it is often misinterpreted by developers. For example, developers should check for \texttt{NULL} values using the \texttt{IS NULL} operator instead of the otherwise syntactically correct \texttt{!= NULL} expression that always returns \texttt{UNKNOWN} in SQL \cite{nagy2017static}.

\textbf{Ambiguous Groups} smell occurs when developers misuse the \texttt{GROUP BY} aggregation command. For example, adding columns in the select list other than the ones used in aggregation function or in \texttt{GROUP BY} clause may generate erroneous results \cite{karwin2010sql}.

\textbf{Random Selection} occurs when developers query a single random row. This operation requires a full scan of the required table. This will have a negative impact on the performance as the size of the table increases \cite{karwin2010sql}.  

SQLInspect supports the detection of these four smells out of the total six types of query smells from the catalog of Karwin; hence, we also rely on these. We notice that the catalogue of Karwin groups smells into the following categories:
Logical Database Design, Physical Database Design, Query,
and Application Development. As our goal is to investigate the application code, relevant ones for us are the last two categories. However, SQLInspect does not implement the detection of smells in the Application Development category as they are not explicitly in the SQL code. %\Foutse{...we have to provide a rational for focusing on these specific ones!!!...is it just because they are the only types of smells that our tool can detected? if so, we should explain it!...in any case we need to give a rationale for picking them!}
SQL Code smell detection in SQLInspect relies on SQL query extraction, which has a minimum precision of 88 \% and a minimum recall of 71.5\% \cite{meurice2016static}. Hence, the aforementioned precision and recall values can be considered as an upper bound for SQL Code smell detection performance. More details on SQLInspect and the supported smells can be found in the related papers of Nagy et al. \cite{nagy2017static, nagy2018sqlinspect}.
%\Foutse{please provide information about the accuracy of the tool used...}. 

\subsection{Apriori: Association Rule Mining Algorithm}
Apriori is an algorithm devised for mining frequent itemsets and relevant association rules \cite{agrawal1994fast}. It has been successfully used to %runs on transactions that contain variables whose co-occurrences are is to be studied. It is used to 
mine association between items in many problems such as market basket analysis \cite{kaur2016market}, intrusion detection \cite{jin2019survey}, supply chain management \cite{agarwal2017decision} and requirement engineering \cite{alzu2018novel}.
The Apriori algorithm first scans the dataset  (i.e., transactions) and generates frequent itemsets based on filtering criteria set by users. Then, a list of association rules is generated from the frequent itemsets.

We use the support \cite{agrawal1993mining}, confidence \cite{agrawal1993mining}, lift \cite{brin1997dynamic}, leverage \cite{piatetsky1991discovery} and conviction \cite{brin1997dynamic} parameters to quantify the degree of association between two items (or smells). The range of values for support and confidence is between 0 and 1. Lift can take any value between 0 and $\infty$. If the value of lift is 1, it means that the smell pairs are independent. Leverage has a range between -1 and 1. A leverage value of zero shows independence. Conviction has a range of 0 and $\infty$. Independent occurrences have a conviction of 1.  

\subsection{Cramer's V Test for Association}

The Cramer's V test measures the level of association between categorical variables \cite{cramir1946mathematical}. It has a value between 0 and 1. A value of 0 indicates complete independence, and a value of 1 indicates complete association. The Cramer's V test takes into account sample size when comparing two variables. The formula is given in Equation \ref{eq:cram} where $X^{2}$ is the Pearson's Chi-square coefficient, $n$ is the total number of samples and $\mathit{row}$ and $\mathit{col}$ represent the number of distinct values of the categorical variables whose association is to be computed. 
%\Foutse{what is a row variable? what is a column variable? what do they represent concretely?}%, respectively.
\begin{equation}
\label{eq:cram}
  V= \sqrt{\frac{X^{2}}{n*\mathit{min}(\mathit{row}-1, \mathit{col}-1)}} 
\end{equation}

\subsection{Survival Analysis}
Survival analysis \cite{miller2011survival} is a statistical analysis technique that provides the expected time of the occurrence of an event of interest. The event of interest could be anything as long as it is clearly defined. We define a study observation window and track events of interest that occur within the window. If the subjects under study leave during the period of observation, the corresponding data will be censored. If the event is not observed during the observation period, the corresponding subject will be censored at the end of the period.  \textit{Time to event} and \textit{status} are two important variables for survival analysis.

\textbf{Time to event (T)} is defined as the time interval between the starting of observation and
the occurrence of an event or the censoring of data.
%from starting of observation  until 
%either an event occurs or the data is censored. 
This time can be measured in any unit. Thus, $T$ is a random variable with positive values \cite{miller2011survival}.
\textbf{Status} is a boolean variable that indicates whether an event is observed or the data is censored. If the event occurs during the observation period, status 
takes a value of 1 and otherwise 0.
%value is one otherwise it will be zero. 
\textbf{The Survival function S(t)} gives the probability ($P(T > t)$) that a subject will survive beyond time $t$. 
After we arrange our data in increasing order of $T$, we can plot the survival curve and estimate the survival probability using one of the commonly used survival estimators (e.g., Kaplan-Meier estimator \cite{kaplan1958nonparametric}). The  Kaplan-Meier estimation is computed following Equation \ref{eq:kapmai}, where $t_{i}$ is the time duration up to event-occurrence point $i$, $d_{i}$ is the number of event occurrences up to $t_{i}$, and $n_{i} $ is the number of subjects that survive just before $t_{i}$. $n_{i}$ and $d_{i}$ are obtained from the aforementioned ordered data. 
\begin{equation}
\label{eq:kapmai}
  S(t)= \prod_{i:t_{i} \leq t}{[1-\frac{d_{i}}{n_{i}}]}
\end{equation}

% -----------------------------------------------------------------------------------------------------------------------------------------------------------------

\section{Study Method}

\begin{figure*}[ht]
    \includegraphics[width=6.5in]{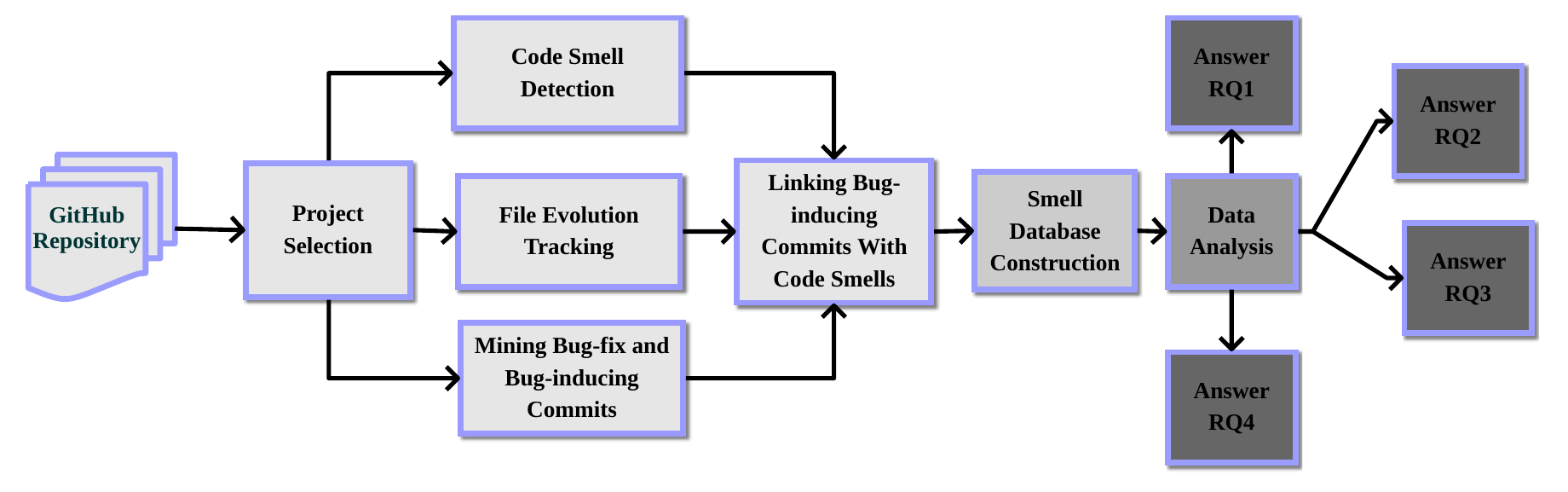}
     \vspace{-.5cm}
    \caption{Process followed to conduct our empirical study}
    \label{fig:Approach}
    \vspace{-.3cm}
\end{figure*}
\label{sec:method}

In this section, we describe our study method where we select appropriate projects for our study, detect traditional and SQL code smells within their source code, extract bug-fixing and bug-inducing commits from their version history, and then analyse all these items to answer our research questions. Figure \ref{fig:Approach} shows the process of our empirical study.
%i and then analyse the
%to collect and analyze smells on selected projects.

\subsection{Project Selection}
%We select software projects from GitHub that are developed in Java, SQLInspect only detects SQL code smells in Java source codes, and have the issue tracking system enabled to filter out personal projects that are not . 
%\Foutse{why these criteria? please provide a justification!!!}. 
 %the existing tools for SQL code smell detection (e.g., 
We limited our study to Java because the SQL code smell detection tool we selected for our study, SQLInspect \cite{nagy2018sqlinspect} can only process programs written in Java. % with Java language code, we only consider Java-based projects from GitHub.
We select our projects from GitHub using four steps as follows.

\textbf{Phase-I:} We use GitHub search mechanism and collect the software repositories labelled with four keywords -- \texttt{android app, hibernate, JPA, Java}. We choose these keywords (a.k.a., categories) since we were interested in data-intensive software systems and also wanted to study SQL code smells in their embedding code.

\textbf{Phase-II:} We performed code search on each project selected in the first phase using GitHub code search API \cite{github_code_search}. In particular, we look for the import statements (e.g., \texttt{import android.database} \texttt{.sqlite.SQLiteDatabase}) that SQLInspect 
can analyze to %might be analysing to 
detect potential SQL code smells.

%was looking for to generate the database smells. 
%For example, import statements such as:
%\textit{android.database\.sqlite\.SQLiteDatabase, org\.hibernate\.Session}. 

\textbf{Phase-III:} Once Phase-I and Phase-II are completed, we collect the projects that (1) fall into the four categories above and (2) pass the constraint of import statements in their source code. 

% \Foutse{what do you mean exactly? why not simply split Phase III in two phases? you do a first round? then you leverage the results to do some kind of snowballing... by extracting interesting key works as described bellow and repeating the search!!! the current presentation is confusing!!!}.
 
\textbf{Phase-IV:} Since the project collection in the above three phases was not significantly high, we thus collect all the labels from each project and build a word-count dictionary to identify the most common keywords. Then we select Top-50 keywords from each of the four categories and repeat Phase-I, which delivers a large collection of 35,000 projects. Then we look for import statements in their source code again, and separate 800 projects that contained the required import statements. 
%\CSABA{Here I don't understand how we end up with only 800 projects if we had 35000 projects before. Is it because only 800 contain import statements we are looking for? yes}

%Finally, we were able to 35,000 projects by iterating on the two phases.

\textbf{Phase-V:} We ran the SQLInspect tool on 800 projects and selected the projects with at least 10 
database access queries. We choose this threshold to capture the projects that vary in size and complexity and  
%\Foutse{what do you mean by reasonable data set? this is vague!!!...} 
obtain a dataset with a significant number of queries for analysis. Finally, we ended up with a total of 150 data-intensive software projects. On average, each project has a size of 146 KLOC, 121 SQL queries and 15 data-access classes. Overall, 13\% of these projects have more than 500 KLOC and 30\% of them use more than 73 SQL queries. About 48\% of SQL queries in those projects perform SELECT operations where 11\% have sub-queries.

%Thus, at least 52 \% of the queries perform more complex operations (e.g., INSERT, UPDATE, JOIN). The largest SQL query involves 49 SUBSELECTs and 32 JOIN operations.}

%cover project of varying size and complexity. Finally, we obtained a 

We also classify our selected projects into four application domains -- \emph{Business, Library, Multimedia} and \emph{Utility} -- to capture the domain-related aspects. 
%for discussion at a more abstract level. 
We assign each project to any of these four groups by consulting their overview on the GitHub pages.  
%to the groups based on information obtained from the projects github page. 
Software projects that are used for business and educational purposes (e.g., data analysis) are kept in the \emph{Business} category. Open-source libraries or tools used by developers are categorized into the \emph{Library} category.
%to develop and maintain software are categorized in to \textbf{Library}. 
Games and media player systems are categorized under \emph{Multimedia}. 
Finally, software projects for personal uses (e.g., task management, scheduling or social networking) 
are categorized into the \emph{Utility} category.

Table \ref{tbl:proj_info} shows, for each application domain, the total number of projects and their median number of database access queries.
%value of the number of database access queries of all projects 
The median is calculated by considering the latest version of all selected projects.  
% Please add the following required packages to your document preamble:
% \usepackage{booktabs}
% \usepackage{graphicx}
\begin{table}[!t]
\centering
\caption{Selected projects \& their database access statistics}
\label{tbl:proj_info}
\vspace{-.2cm}
\resizebox{.80\linewidth}{!}{%
\begin{threeparttable}
\begin{tabular}{@{}lcc@{}}
\toprule
\textbf{Application domain} & \textbf{\#Projects} & \textbf{Median DAQC} \\ \midrule
Library & 97 & 32 \\
Business & 23 & 46 \\
Utility & 19 & 50.5 \\
Multimedia & 11 & 19.5 \\ 
\bottomrule
\end{tabular}%
\centering
\textbf{DAQC} = Database Access Query Count
\end{threeparttable}
}
\vspace{-0.4cm}
\end{table}
\subsection{Code Smell Detection}
It is not practical to detect code smells from every commit of each project due to the large number of projects and commits. Therefore, we detect the traditional and SQL code smells from each project by taking their snapshots after every 500 commits starting from the most recent commits backwards. A similar approach was followed by Aniche et al. \cite{aniche2018code}. 

%\Foutse{please add reference to the paper that studied code smells in Model view control architetcures!!!}

We use SQLInspect \cite{nagy2018sqlinspect}, a static analysis tool, for SQL code smell detection. SQLInspect extracts SQL queries from the Java code and then detects four types of SQL code smells -- Implicit Columns, Fear of the Unknown, Random Selection and Ambiguous Groups. The tool can detect smells from the SQL code targeting several database access frameworks -- \textit{Android Database API}, \textit{JDBC}, \textit{JPA}, and \textit{Hibernate}. 

%\Foutse{if not mentioned before...you should say something about the accuracy of thus tool!!!}

%SQLInspect can identify queries and smells for projects that use SQLLite, JDBC, JPA and hibernate frameworks for data access. 

We use DECOR \cite{gueheneuc2007ptidej}, a reverse engineering tool, for detecting the traditional code smells. DECOR can detect 18 different traditional code smells from Java source code. DECOR has a recall of 100\% and a precision > 60\% \cite{moha2010domain}. %precision on smell detection.  
%\Foutse{similarly...you should say something about the accuracy of thus tool!!!}

\subsection{Tracking Project File Evolution}
Software projects change and so are their source code files 
%the location and name of files in a project 
as they evolve over time. To ensure a reliable analysis of software evolution, file genealogy tracking is important. 
Tracking of file status can help us resolve the issues involving file renaming or file location changes during evolution.
%helps to ensure that similar files with different names as a result of renaming and location change will be treated as one. 
We use the \texttt{git diff} command to compare two consecutive project snapshots using their commit identifiers. The command
%Git diff with the option of name-status 
shows a list of files that are either added, deleted, modified or renamed between two given commits.
It also provides a numerical estimation on how likely a file has been renamed. We consider a threshold of 70\% accuracy to detect file renaming, as was used by an earlier study \cite{johannes2019large}. 
%accuracy of rename detection as a percentage value. We considered renames with accuracy greater or equal to 70\% as true renames. 
%70\% threshold is also used in the work of David Johannes et al. \cite{johannes2019large}. 
Finally, each source file in each of our projects is tagged with a unique identifier generated from the file tracking information.

\subsection{Mining Bug-Fix and Bug-Inducing Commits}
We use PyDriller \cite{PyDriller} to mine bug-fixing and bug-inducing commits from our selected projects. PyDriller offers a Python API that interacts with any GitHub repository using a set of Git commands.

To identify bug-fix commits using PyDriller, we employed a set of 57 keywords that indicate possible fixing of bugs, errors and software failures (e.g., \textbf{fix, fixed, fixes, bug, error, except, issue, fail, failure, crash}). The set of keywords were selected based on the work of Mockus and Votta \cite{mockus2000identifying} and Antoniol et al. \cite{antoniol2008bug}, who showed that those keywords have a tendency to be associated with bug-fix commits. These keywords were also used in multiple previous studies to identify bug-fixing commits \cite{kamei2012large,kim2008classifying, guerrouj2017investigating}. The complete keyword list is available in the replication package \cite{replication}. Our tool searches for each keyword in the commit messages, and separates the commits containing the keywords as bug-fixing commits.  

%\Foutse{how accurate is this heuristic? we should say something about it or explain that it has been used in multiple previous studies and discuss issues related to its accuracy in the threats to validity section!!!}
%tracked the first keyword associated with a commit. Table 
Table \ref{tbl:bug_fix_keywords} shows the proportion of bug-fix commits that are identified using the top six prevalent keywords.

PyDriller implements the SZZ algorithm \cite{sliwerski2005changes} to pinpoint a bug-inducing commit from a given bug-fix commit within the version-control history. We use PyDriller to detect the bug-inducing commits for the bug-fixing commits detected above. 
%We used this API for all identified bug fixing commits and obtained the corresponding bug inducing commits. We store commit date, commit message, author information and the changed source files from each of the bug-inducing commits for our analysis.

%\Foutse{don't forget to mention the limitations of this SZZ algorithm in the threats to validity section!!!}
%the name of all touched files on bug-inducing commits. 

\subsection{Linking Bug-Inducing Commits with Code Smells}

To determine any association between code smells and software bugs, the smells have to be present in the code before the bugs actually occur.
%For a association between smell and bug, the smell has to occur before the bug. 
We determine such potential causal associations using bug-inducing commits.
%associated the bug inducing commits and smells by maintaining causality. 
Let $T_{0}$ be the snapshot date of the smelly code file 
and $T_{n} $ be the commit date of the next snapshot that tracks the same code file.
Now, we identify the bug-inducing commits between $T_{o}$ and $T_{n}$ that contain the smelly code file from 
%with this file at the 
version $T_{o}$. If any bug-inducing commit touches the smelly file which is later fixed in the corresponding bug-fixing commit, then 
we mark such smells as linked with the target bug-inducing commits.

\begin{table}[!t]
\centering
\caption{Most prevalent keywords used to detect bug-fix commits}
\label{tbl:bug_fix_keywords}
\vspace{-.2cm}
\resizebox{2.5in}{!}{%
\begin{tabular}{@{}lc@{}}
\toprule
\textbf{Keywords} & \textbf{Bug-Fix Commits} \\ \midrule
fix, fixed, fixes & 66.16\% \\
bug & 7.93\% \\
issue & 6.16\% \\
except & 4.84\% \\
error & 4.51\% \\
fail, failure & 3.55\% \\
\bottomrule
\textbf{Total bug-fix commits} & \textbf{110,747} \\ \bottomrule
\end{tabular}%
}
\vspace{-.434cm}
\end{table}

\subsection{Construction of a Smell Database}
In order to perform our analysis reliably, we store the information extracted from the earlier steps in a relational database. 
%Since our analysis is at a class level. Our dataset is organized to work with files and snapshot versions. 
A record in the smells table of our database is identified using a combination of file identifier and project version number (a.k.a., file-version-ID). Each record comprises of 
%by file id and version number together. We use \textbf{file-version} for this and subsequent discussion to denote the combination of file id and version as an identifier of a row. 
%For each file-version, we have 
a vector that stores the statistics on traditional code smells, SQL code smells found within a source code file and its bug-inducing related meta data. Our database contains a total of 1,077,548 records for 139,017 source files from 150 projects with 1648 versions. 
%\textbf{However, our study focuses on SQL code smells that can only occur in file-versions that access database, we only used file-versions that have at least one SQL query to answer all research questions.}
However, our study analyzes only such records where the source code files deal with database access, and might contain SQL code smells. Thus, in practice, we deal with a subset of 29,373 records for our study. 

%contain both SQL code smells and traditional code smells.

%code  found in the corresponding source code file.
%to store the number of database accesses and the corresponding database smell count and traditional smell count. In addition we have "in\_bug\_induce" column to store a Boolean value if a file-version is mentioned in bug-inducing commits. 

%We only considered file-versions that have at least one database access query to answer RQ2, RQ3, and RQ4 as database smells can only occur in file-versions that access database. We also transformed the smell counts to 1 (present) and 0 (absent). This operation is required for RQ2, RQ3, and RQ4 analysis but not for RQ1.

%from database smells and \ from traditional code smells. The four smells are selected as they are most prevalent in our data.

\subsection{Experimental Data Analysis}

\textbf{Association between SQL and Traditional Code Smells:} For Apriori analysis, we consider each entry (i.e., a record from our database that has at least one database access query) containing code smell statistics as a transaction. Then, the frequent itemsets are generated from all the transactions that involve traditional code smells and SQL code smells. 

%In particular, we select two most prevalent traditional smells (e.g., LongMethod and LongParameterList) and two most prevalent SQL code smells (e.g., Implicit Columns and Fear of the Unknown) for our association analysis.

Besides the Apriori algorithm, we employ Cramer’s V association test to collect numerical, comparable association values between these two classes of code smells (RQ2).
%To investigate the Co-occurrence between database smells
%and traditional code smells we run the Apriori Association rule mining algorithm on our dataset. We also run 

\textbf{Co-occurrence between SQL Code Smells and Bugs:}
To investigate the co-occurrence (or potential causation) between SQL code smells and software bugs (RQ3), we employ both Chi-squared test and Cramer's V test. We also 
%to investigate the association between smells and bugs. In addition, we built a
develop a RandomForest model to investigate the importance of various code smells in determining %whether a given commit is bug-inducing or not.
co-occurrence with bugs.

%observe the contribution of smells to classify file-versions as bug-inducing or not.

\textbf{Survival Analysis of Code Smells:} We analyze the survival rates of 
traditional and SQL code smells during the evolution of our selected systems (RQ4). For survival analysis, we use 
the Kaplan-Meier curve \cite{kaplan1958nonparametric} (e.g., Fig. \ref{fig:RQ4_imp_col}). The curve shows the survival probability $S(t)$ of a given code smell at a time $t$. 
We define the fixing of a code smells as our \emph{event of interest}. That is, if a source code file contains a target smell in an earlier version snapshot and does not contain the same smell in the current snapshot, our event of interest occurs at the current snapshot. The occurrence of this event determines the survival probability of corresponding code smell.   

%our event of interest  \textbf{Smell is fixed} and \textbf{status} as follows.
 %\textbf{Smell is fixed} event occurs when a file that already has a smell in previous versions does not have a smell at the current version. \textbf{Status} variable indicates if a file-version is censored or not. If \textbf{smell is fixed} event happens any time in the observation period, \textbf{status} will be one. Otherwise, status will be zero, indicating the data is censored. Our observation period covers the time from just before the oldest snapshot to the time after the newest snapshot to include all snapshots in our analysis.
 
\subsection{Replication Package}

We made our collected data and results publicly available \cite{replication}. We provide (i) the database of smells in the projects under question, (ii) the list of keywords used for the identification of bug-fixing commits, (iii) data collection and analysis scripts.

% -----------------------------------------------------------------------------------------------------------------------------------------------------------------

\section{Study Results}
\label{sec:find}

In this section, we present our study findings and answer each of the four research questions as follows.

\subsection{RQ1: What is the Prevalence of SQL Code Smells Across Different Application Domains?}

\begin{figure}[!htb]
    \includegraphics[width=2.5in]{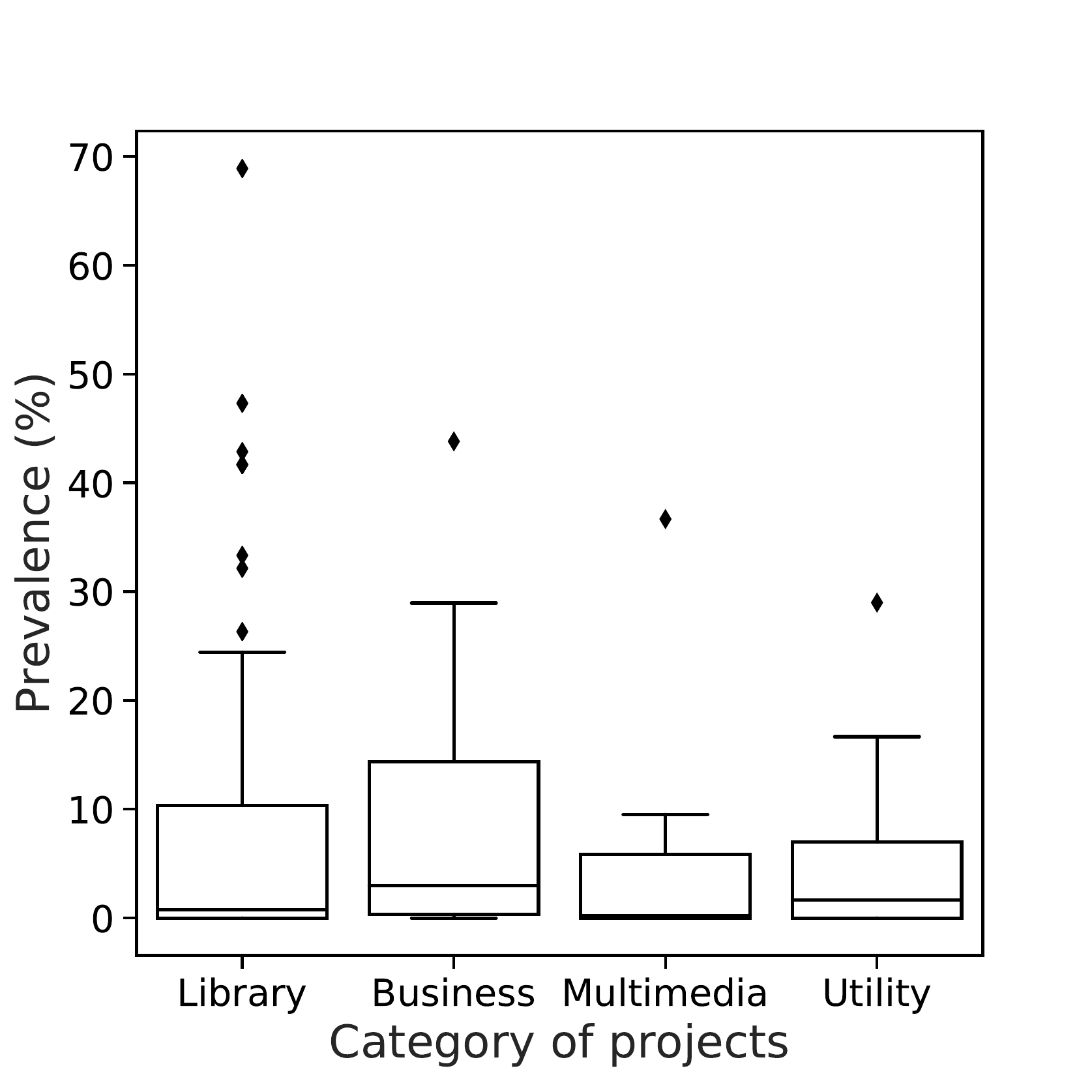}
    \caption{Prevalence of SQL code smells (Implicit Columns) across different application domains}
    \label{fig:pre_impl}
\vspace{-0.5cm}
\end{figure}

\begin{figure}[!htb]
    \centering
    \includegraphics[width=2.8in]{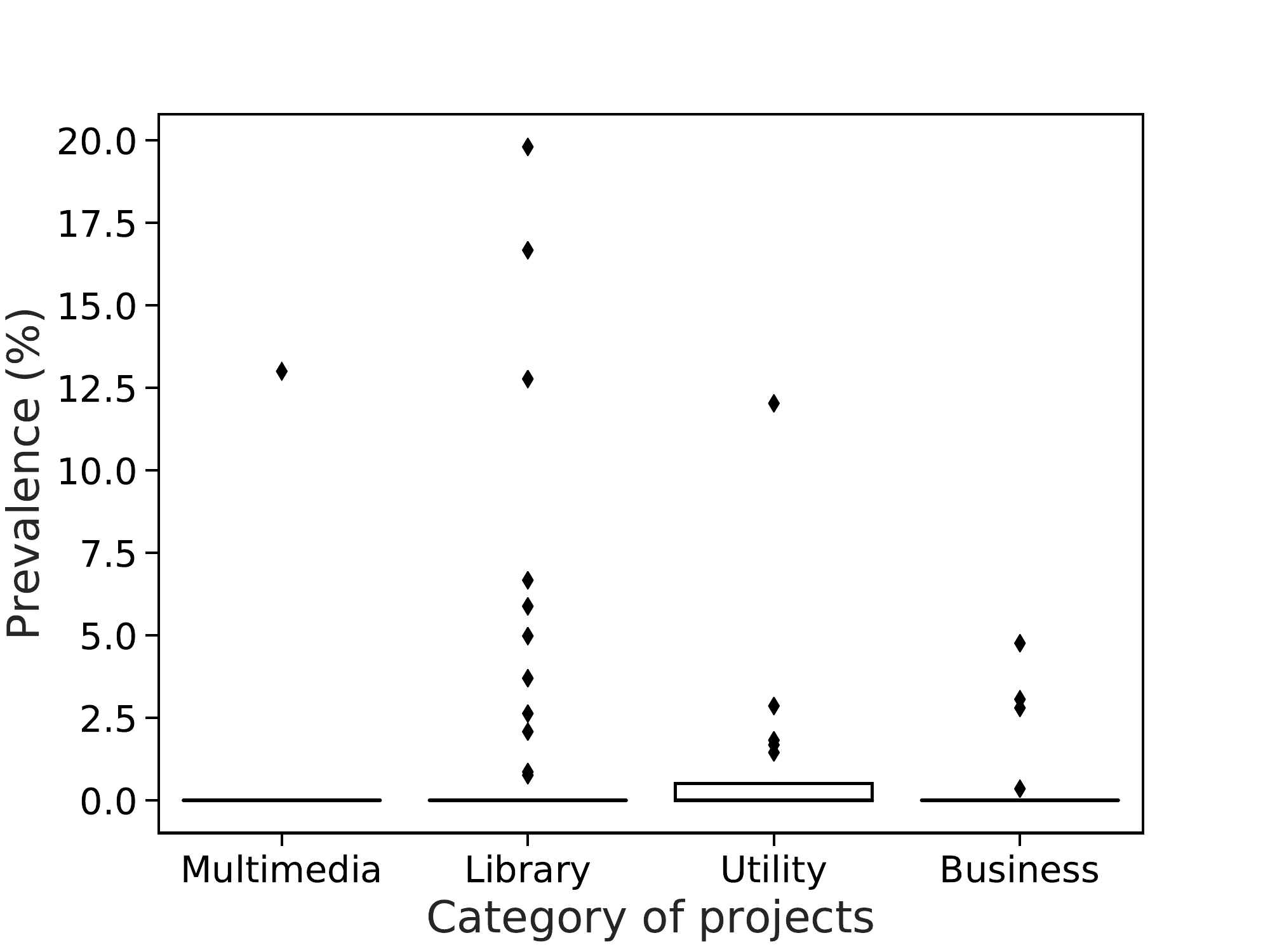}
    \vspace{-.2cm}
    \caption{Prevalence of SQL code smells (Fear of the Unknown) across different application domains}
    \vspace{-.4cm}
    \label{fig:pre_fear_app}
    
\end{figure}

We collect SQL code smells from each of the projects (i.e., latest tracked versions) and provide the summary statistics on code smells for each of the four application domains.
%summarize SQL code smell statistics of all projects using the projects' latest tracked version and by grouping each project by its application domain. 
Our projects from each domain have varying complexity in terms of project size and interactions with the database. 
We determine \textit{prevalence} of SQL code smells as the ratio between total number of SQL code smells and total number of database access queries in a subject system. We present our prevalence analysis in Figures \ref{fig:pre_impl}, \ref{fig:pre_fear_app} and Table \ref{tbl:app_implicit}.

\begin{table}[!ht]
\caption{Prevalence of Implicit Columns across four application domains}
\resizebox{3in}{!}{%
 \begin{tabular}{@{}lcc@{}}
\toprule
\textbf{Domain} & \textbf{Median Prevalence} & \textbf{Mean Prevalence} \\ \midrule
Business & 2.98\% & 8.49\% \\
Multimedia & 0.23\% & 5.47\% \\
Utility & 1.68\% & 5.27\% \\
Library & 0.75\% & 7.93\% \\ 
\bottomrule
\end{tabular}%
}
\label{tbl:app_implicit}
\vspace{-0.7cm}
\end{table}

\vspace{0.4cm}
We detect four types of SQL code smells (e.g., Section \ref{sec:sql-smells}) with SQLInspect in our data-intensive subject systems. Out of these four smell types,  \emph{Implicit Columns} was the most frequent across all projects with a median prevalence of 1.67\%. That is, out of every 100 database access queries, two queries are affected by this smell. The second most frequent code smell -- 
%(median prevalence= 1.67\%) is the most commonly occurring across all projects, the second most frequent database smell is 
\textit{Fear of the Unknown} -- has a median prevalence of 0.8\%.  
%database smell (median prevalence=0\%, mean prevalence = 0.8\%). 
We did not find any \textit{Ambiguous Groups} or \textit{Random Selection} in the most recent tracked version of our subject systems. However, our analysis identified a few \textit{Ambiguous Groups} code smells in the older versions of the systems.

We analyse the prevalence of \emph{Implicit Columns} across four application domains. Table \ref{tbl:app_implicit} and Fig. \ref{fig:pre_impl} summarize our findings. From Table \ref{tbl:app_implicit}, we see that projects from
%prevalence of all project categories. 
\emph{Business} and \emph{Utility} domains have the highest median prevalence of 2.98\% and 1.68\%. Fig. \ref{fig:pre_impl} further shows the distribution of prevalence for \emph{Implicit Columns} across the application domains. We see that \emph{Business} and \emph{Library} have the highest median and 75\% quantile in the prevalence ratio measure. We also investigated the nature of the outlier projects from Library domains, as shown in the box plot of Fig. \ref{fig:pre_impl}. We notice that the Library project with the highest prevalence ratio, \textit{Tablesaw} data visualization library, has 45 SQL queries, out of which 31 queries are smelly. This project has more than 2K stars and 39 contributors. The second highest in prevalence, \textit{calcite-elasticsearch}, is another library project that has a total of 167 SQL queries, out of which 79 queries are smelly. Since library projects are often reused by other applications, the impact of these SQL code smells could be much more serious.
%The fact that both projects are library projects is more concerning as the impact of smells could propagate to other software that is dependent on the library projects.
In both Figure \ref{fig:pre_impl} and Table \ref{tbl:app_implicit}, we see that SQL code smells such as \emph{Implicit Columns} have the least prevalence in the subject systems from \emph{Multimedia} domain.

%ratio of \textit{Implicit Columns} smell from all groups. Multimedia applications have the lowest median ratio.  
%Figure \ref{fig:pre_impl} and Table \ref{tbl:app_implicit} summarize the prevalence of \textit{Implicit Columns} smells on the four category of projects.

We further analyse the distribution of prevalence for \emph{Fear of the Unknown} SQL code smell across the four application domains. Fig. \ref{fig:pre_fear_app} shows 
our prevalence ratio distribution for this smell. 
%of fear of the unknown database smell across all application domains. 
We see that the median prevalence for all domains is zero. However, there exist a significant number of outlier projects in the library, business and utility domains that we analyse. The project with the highest prevalence of \emph{Fear of the Unknown} smell is a real-time chat and messaging Android SDK library, \textit{Applozic-Android-SDK}, that has at least 295 forks and 18 contributors. The project has 202 SQL queries in the most recent tracked version, out of which 40 queries are affected with the target smell.

 All our analyses above show that \textit{Implicit Columns} and \textit{Fear of the Unknown} are the two prevailing SQL code smells across all four application domains. 
 %\textit{Implicit Columns} smells are often introduced by the developers when they attempt to write SQL queries quickly without mentioning the expected column names. 
We also randomly selected 10 projects and manually investigated 98 \textit{Implicit Columns} smells from them. We found that at least 70\% of these smelly SQL queries retrieved three or more columns that were unused and 15\% retrieved nine or more table columns that were unused.
Such a counter-productive data access could lead to a performance bottleneck.
 %However, \textit{Implicit Columns} smells have an impact on maintenance as well as performance. 
 \textit{Implicit Columns} smells might also create unnecessary coupling between a front-end and its back-end database, which could negatively affect the maintainability of the system. 
 Although the prevalence of \textit{Fear of the Unknown} smell is not as high as for \textit{Implicit Columns}, their impact on maintenance and performance could also not be ignored. 
 
 %In particular, this smell could lead to logical errors that may skip testing and accidentally surface later in the production stage.

\begin{tcolorbox}[colback=white, colframe=black,left=2pt,right=2pt,top=1pt,bottom=1pt]
\textbf{\textit{Implicit Columns} smells are the most prevalent SQL code smells in the data-intensive systems across four application domains followed by the \textit{Fear of the Unknown} smells. The remaining two SQL code smells are not prevalent in the 150 subject systems under our study.
%across all project groups followed by \textit{fear of the unknown} smells considering the most recent tracked versions of all projects.
}
\end{tcolorbox}

\subsection{RQ2: Do Traditional Code Smells and SQL Code Smells Co-occur at Class Level?}

We determine co-occurrences between SQL code smells and traditional code smells within our subject systems where we consider multiple versions of the source code files (a.k.a., revisions). Table \ref{tbl:rq2_dataset} shows the statistics on file versions for each application domain.  We see that business systems have the highest number of file versions that deal with database access 
while multimedia systems have the lowest number. 
Business systems have more database interactions since they are often involved in data processing and data visualization. We have only 11 Multimedia systems in our dataset, which might explain their low number.   

\begin{table}[!ht]
\caption{Source code file versions with database access}
\label{tbl:rq2_dataset}
\vspace{-.3cm}
\resizebox{2.2in}{!}{%
\begin{tabular}{@{}lr@{}}
\toprule
\textbf{Application Domain} & \textbf{\# File Versions} \\ \midrule
Business & 16,225 \\
Library & 11,839 \\
Multimedia & 156 \\
Utility & 1,153 \\ \bottomrule
\end{tabular}
}
\end{table}

%Although library projects are larger, the amount of data access is lower than business applications. 

%\begin{table}[!ht]
%\caption{Input Parameters for Apriori Algorithm}
%\label{tab:my-table}
%\resizebox{2.6in}{!}{%
%\begin{tabular}{@{}ll@{}}
%\toprule
%\textbf{Parameter} & \textbf{Value} \\ \midrule
%Minimum\_support & 0.01 \\
%Maximum\_length & 2 \\
%Metric for selecting association rules   & Lift \\
%Minimum lift threshold & 1 \\ \bottomrule
%\end{tabular}
%\label{tbl:apriori_params}
%}

%\vspace{-0.2cm}
%\end{table}

We use Apriori algorithm for determining the association (co-occurrence) between traditional code smells and SQL code smells. 
%The used parameters by the algorithm
%The input parameters of the Apriori algorithm are listed can be found in Table \ref{tbl:apriori_params}. 
To generate frequent itemsets, we selected a minimum support of 0.01 (1\%) considering the small number of occurrences of SQL code smells compared to that of traditional code smells. We also restrict the maximum number of items in every itemset to 2 since we were interested in the association between one traditional smell and one SQL code smell. We also set the minimum lift threshold to 1 to generate the \emph{relevant} association between SQL code smells and traditional code smells.

Table \ref{tbl:RQ2_association_rules} shows our frequent itemsets where each itemset comprises of one traditional code smell and one SQL code smell.  When all subject systems are considered, we see an association (i.e., Lift$>1.00$) between  \textit{Implicit Columns} and  \emph{LongMethod}. We also repeat the same experiments for each of the four application domains. We see that \textit{Implicit Columns} smells co-occur with \emph{LongMethod} across both business and library domains.
%generated smell pairs and their corresponding association metric. On the combined dataset, only the smell pair \textit{Implicit Columns: LongMethod} has a Lift greater than 1. We also run Apriori for each application domain separately.
%\textit{Implicit Columns: LongMethod} appeared in business as well as library projects. 
They also co-occur with \textit{ComplexClass} in all application domains except business. However, the leverage value is close to zero for each of the mined association rules, which indicates that the association between SQL code smells and traditional code smells is not strong. 

\begin{table*}[!ht]
\caption{Top-3 SQL code smells and traditional code smells based on lift value 
across the application domains. A leverage value close to 0 indicates weak association.
%for the combined data and each application domain.
}
\vspace{-.2cm}
\label{tbl:RQ2_association_rules}
\resizebox{\textwidth}{!}{%
\begin{tabular}{@{}lllllll@{}}
\toprule
\textbf{Application Domain} & \textbf{Smell Pais} & \textbf{Support} & \textbf{Confidence} & \textbf{Lift} & \textbf{Leverage} & \textbf{Conviction} \\ \midrule
\textbf{Combined} & Implicit Columns:LongMethod & 0.0507 & 0.528 & 1.03 & \textbf{0.0015} & 1.0336 \\ \midrule
Business & Implicit Columns:ComplexClass & 0.0169 & 0.445 & 1.2169 & \textbf{0.003} & 1.1429 \\
 & Implicit Columns:LongMethod & 0.0207 & 0.5437 & 1.031 & \textbf{0.0006} & 1.0358 \\
 \midrule
Library & Implicit Columns:LongParameterList & 0.0295 & 0.1804 & 1.0261 & \textbf{0.0007} & 1.0056 \\
 & Implicit Columns:LongMethod & 0.0854 & 0.5228 & 1.0377 & \textbf{0.0031} & 1.0398\\
 \midrule
Multimedia &Fear of the Unknown:AntiSingleton & 0.01923 &0.2143 &5.5714&	\textbf{0.0158}&1.2238
 \\
 & Fear of the Unknown:ComplexClass & 0.0705 &	0.7857 & 2.7238 &	\textbf{0.0446} & 3.32
 \\
 & Implicit Columns:ComplexClass & 0.1474&0.5476&1.8984&\textbf{0.0698}&1.5729
 \\
 \midrule
Utility & Fear of the Unknown:AntiSingleton & 0.0208&0.31169&4.6074&	\textbf{0.0163}&1.3545
 \\
 & Fear of the Unknown:LongParameterList & 0.01908&0.2857&1.8&\textbf{0.0085}&	1.1778
 \\
 & Implicit Columns:ComplexClass & 0.0928&0.4693&1.5593&\textbf{0.0333}&1.3173
 \\ \midrule
\end{tabular}
}
\vspace{-6pt}
\end{table*}

\begin{table*}[!ht]
\centering
\caption{Chi-square and Cramer's V value of smell pairs computed on the combined dataset for each smell pair in Table \ref{tbl:RQ2_association_rules}. We
reject $H_{0}$ for all smell pairs in bold.}
\label{tbl:RQ2_stat_test}
\resizebox{4.8in}{!}{%
\begin{tabular}{@{}lcl@{}}
\toprule
\textbf{Smell Pairs} & \textbf{Chi-square P-value} & \textbf{Cramer's V} \\ \midrule
\textbf{Implicit Columns:LongParameterList} & \textbf{< 0.0001} & \textbf{0.0708} \\
\textbf{Fear of the Unknown:LongMethod} & \textbf{< 0.0001} & \textbf{0.048} \\
\textbf{Fear of the Unknown:LongParameterList} & \textbf{< 0.0001} & \textbf{0.03864} \\
\textbf{Implicit Columns:ComplexClass} & \textbf{< 0.0001} & \textbf{0.02925} \\
\textbf{Implicit Columns:AntiSingleton} & \textbf{< 0.0001} & \textbf{0.0282} \\
\midrule
\textbf{Fear of the Unknown:ComplexClass} & \textbf{0.02217} & 0.01335 \\
\textbf{Fear of the Unknown:AntiSingleton} & \textbf{0.04868} & 0.0115 \\
Implicit Columns:LongMethod & 0.0796 & 0.01 \\ \bottomrule
\end{tabular}%
}
\vspace{-.2cm}
\end{table*}

We also conduct Chi-squared and Cramer's V tests to check whether the associations between traditional code smells and SQL code smells (e.g., Table \ref{tbl:RQ2_association_rules}) are statistically significant or not.
%and to measure the degree of association for each pair in Table . 
Table \ref{tbl:RQ2_stat_test} shows the p-values from our Chi-squared tests. We assume this null hypothesis -- $ H_{0} $: traditional code smells and SQL code smells occur independently. However,  
%and using $\alpha=0.01$. The result shows that 
given the p-values ($<0.05$) in Table \ref{tbl:RQ2_stat_test}, we have strong evidence to reject the null hypothesis for each of the five emboldened smell pairs.
That is, \emph{Implicit Columns} has a significant association with several traditional code smells such as LongParameterList and ComplexClass. It should be noted that each of these code smells is a result of bad programming practices by the developers.
%for all smell pairs indicated in Bold. 
%\Foutse{isnt this next sentence a bit contradictory? are you sure it is the threshold $<=0.05$ that you meant? or is it $ >=0.05$ and are you referring to all smells or just the non significant ones?}
Given the p-values ($>=0.05$) in Table \ref{tbl:RQ2_stat_test}, we have weak evidence to reject the null hypothesis, i.e., such code smell pairs might not be associated. 

%$ H_{0} $ for the rest of the smell pairs as their P-value is not far from 0.01.

We also further investigate the statistically significant associations between traditional and SQL code smells within our subject systems, and determine the degree of associations using Cramer's V tests. Table \ref{tbl:RQ2_stat_test} shows the results from these tests. We see that 
%, we can not compare the degree of association using Chi-squared test. Cramer's V test provides comparable association values. 
\textit{Implicit Columns:LongParameterList} pair has the highest degree of association with a $V$ value of 0.07, which is still a weak association. 
The smell pairs for which we accept the null hypothesis have also small Cramer's V values, which is expected.

\begin{tcolorbox}[colback=white, colframe=black,left=2pt,right=2pt,top=1pt,bottom=1pt]
\textbf{Several traditional code smells (e.g., LongParameterList) and SQL code smells (e.g., Implicit Columns) could co-occur within the data-intensive subject systems. However, their association is rather weak according to multiple statistical tests and our extensive analysis.}
\end{tcolorbox}
  \vspace{-8pt}
\subsection{RQ3: Do the SQL Code Smells Co-occur with Software Bugs?}
We determine the association between SQL code smells and software bugs by analysing smelly code, bug-fixing code, and bug-inducing code. Our dataset contains
a total of 21,973 file revisions, out of which 3,215 revisions were found in the bug-inducing commits. It should be noted that bug-inducing commits lead to software bugs, which are confirmed by the bug-fixing commits later.
We thus separate the bug-inducing commits, and determine the pair-wise co-occurrence (association) between SQL code smells and bugs within these commits.
We conduct Chi-squared test and Cramer's V test to check the significance and degree of the association. Table \ref{tbl:rq3_chi_square} shows our investigation details.
%are mentioned in bug inducing commits. 
%We tested the pairwise co-occurrence of smells with bug inducing column using Chi-squared test and Cramer's V test. 

%the file-version being bug-inducing or not.
To determine the association between SQL code smells and software bugs, we assume this null hypothesis -- $ H_{0} $: The presence of SQL code smells in a file version and the file version being bug-inducing
%SQL code smells and the file-version being bug inducing 
are independent phenomena.
We test this hypothesis with Chi-squared test using $\alpha=0.05$. As shown in Table 
%. We report the result in Table 
\ref{tbl:rq3_chi_square},  we notice that both \textit{Implicit Columns} and \textit{Fear of the Unknown} are two SQL code smells that occur independently of the bug-inducing commits. They have p-values greater than our significance threshold of $0.05$. 

%However, \textit{Fear of the Unknown} has a lower p-value than that of \textit{Implicit Columns}\Foutse{having a lower p-value doesnt mean much, you should compute odd ratio instead or any other effect size metric if you want to compare the strenghts of two associations!!! you can use Cramer's V value to compare...but not p-values!}.
Although the \textit{Implicit Columns} smell is known to cause performance issues and software bugs \cite{karwin2010sql,nagy2017static}, our empirical analysis did not show a strong correlation with bugs.
%\textit{Fear of the Unknown} smell may lead to some exceptions or bugs.
On the contrary, the traditional code smells such as 
%compared to \textit{IMPLICIT \_COLUMNS} which implies that it has more association with bugs. traditional code smells, 
\textit{SpaghettiCode, ComplexClass} and \emph{AntiSingleton} have significant p-values $<0.05$, which indicates that they have a stronger association %evidence of co-occurrence 
with bugs. The traditional code smell namely \textit{ComplexClass} has the lowest and the most significant p-value, which indicates its significant association with the bugs. \emph{ComplexClass} was also reported to be associated with software bugs by the earlier studies \cite{li2007empirical,zazworka2011investigating}.

%The impact of \textit{Implicit Columns} on bugs is small as it is more associated with performance issues and not necessarily bugs \cite{nagy2017static}. 

%To compare smells considering effect of sample size, 
We also determine the degree of association between any code smells and software bugs using Cramer's V test.
%measured the association of database smells and traditional code smells using . 
As shown in Table \ref{tbl:rq3_chi_square}, we see that the traditional code smells (e.g., \textit{LongMethod}, \emph{LongParameterList}) have a relatively higher V-values than that of SQL code smells. That is, SQL code smells might be less associated with the bugs than the traditional code smells.

%shows that With the exception of, traditional code smells have higher association with bugs compared to database smells.

We also develop a RandomForest model to investigate the contribution of each code smell on determining whether a file revision is bug-inducing (e.g., true class) or not (e.g., false class). Since the dataset was not balanced, we used  SMOTE-based oversampling \cite{chawla2002smote} and 10-fold cross-validation for our machine learning model. Finally, we collect the feature importance values from our trained model. These values indicate the importance of code smells (i.e., predictors) on determining whether a file version being bug-inducing or not.
%using RQ3 dataset to investigate the contribution of each smell as a predictor to the file-version being bug inducing or not. 
%We balanced the number of bug-inducing classes in the dataset synthetic minority over-sampling technique . We generated 10 different random forest models using 10-Fold cross validation to account for model variation. We set the number of estimators to 100 in all cases. 
The last column of Table  \ref{tbl:rq3_chi_square} shows how each of the code smells could turn its containing file to be bug-inducing. 
%result of random forest model trained on RQ3 dataset. 
We see that \textit{ComplexClass} has the highest importance of 46\%. Despite the low Cramer's V values, \textit{LongMethod} and \emph{LongParameterList} 
are pretty important ($\approx$10\%) in our trained model.
%have the next highest contribution. While Cramer's V computes pairwise association, random forest model considers the combined impact of smells which explains the difference. 
On the other hand, SQL code smells (e.g., Fear of the Unknown, Implicit Columns) might be less important according to our model, which clearly indicates their low association with the software bugs.  
%\textit{Implicit Columns} smell have the lowest contribution, among the smells in Figure \ref{fig:RQ3_feat_imp}, to bugs. 

\begin{table}[]
\centering
%\caption{Chi-squared P-value and Crammer’s V result of association between smells and buggy files}
\caption{Result of statistical tests and random forest model of association between smells and buggy files.}
\label{tbl:rq3_chi_square}
\vspace{-.2cm}
\resizebox{3.3in}{!}{%-
\begin{tabular}{@{}llll@{}}
\toprule
\textbf{Smell} & \textbf{Chi-square} & \textbf{Cramer's} & \textbf{Feature} \\ 
 & \textbf{P-value} & \textbf{V value} & \textbf{Contribution (\%)} \\
\midrule
Implicit Columns & 0.2377 & 0.0069 & 5.095 \\
Fear of the Unknown & 0.1671 & 0.008 & 7.695 \\
LongMethod & 0.1162 & 0.0003 & 9.86 \\
\textbf{LongParameterList} & \textbf{0.0034} & 0.0034 & 9.1 \\
\textbf{AntiSingleton} & \textbf{< 0.001} & 0.0001 & 7.69 \\
\textbf{SpaghettiCode} & \textbf{< 0.001} & 0.0227 & 5.87 \\
\textbf{ComplexClass} & \textbf{< 0.001} & 0.0846 & \textbf{46.65} \\ \bottomrule
\end{tabular}%
}
\vspace{-0.5cm}
\end{table}

\begin{tcolorbox}[colback=white, colframe=black,left=2pt,right=2pt,top=1pt,bottom=1pt]
\textbf{SQL code smells (e.g., Implicit Columns, Fear of the Unknown) do not have statistically significant association with software bugs. On the contrary, traditional code smells (e.g., ComplexClass, SpaghettiCode) have a statistically significant association with the bugs, according to the results of the performed two statistical tests and RandomForest-based feature contribution analysis.}

\end{tcolorbox}
  \vspace{-8pt}
%\textbf{\textit{Fear of the Unknown} database smell has a statistically significant co-occurrence with bugs. However, Cramer's V test shows small association value with bugs. \textit{Implicit Columns} database smell has even smaller association. traditional code smells have more association with bugs compared to database smells.} 

\subsection{RQ4: How Long do the SQL Code Smells Survive?}

\begin{figure}[!ht]
%    \vspace{-6pt}
    \includegraphics[width=2.4in]{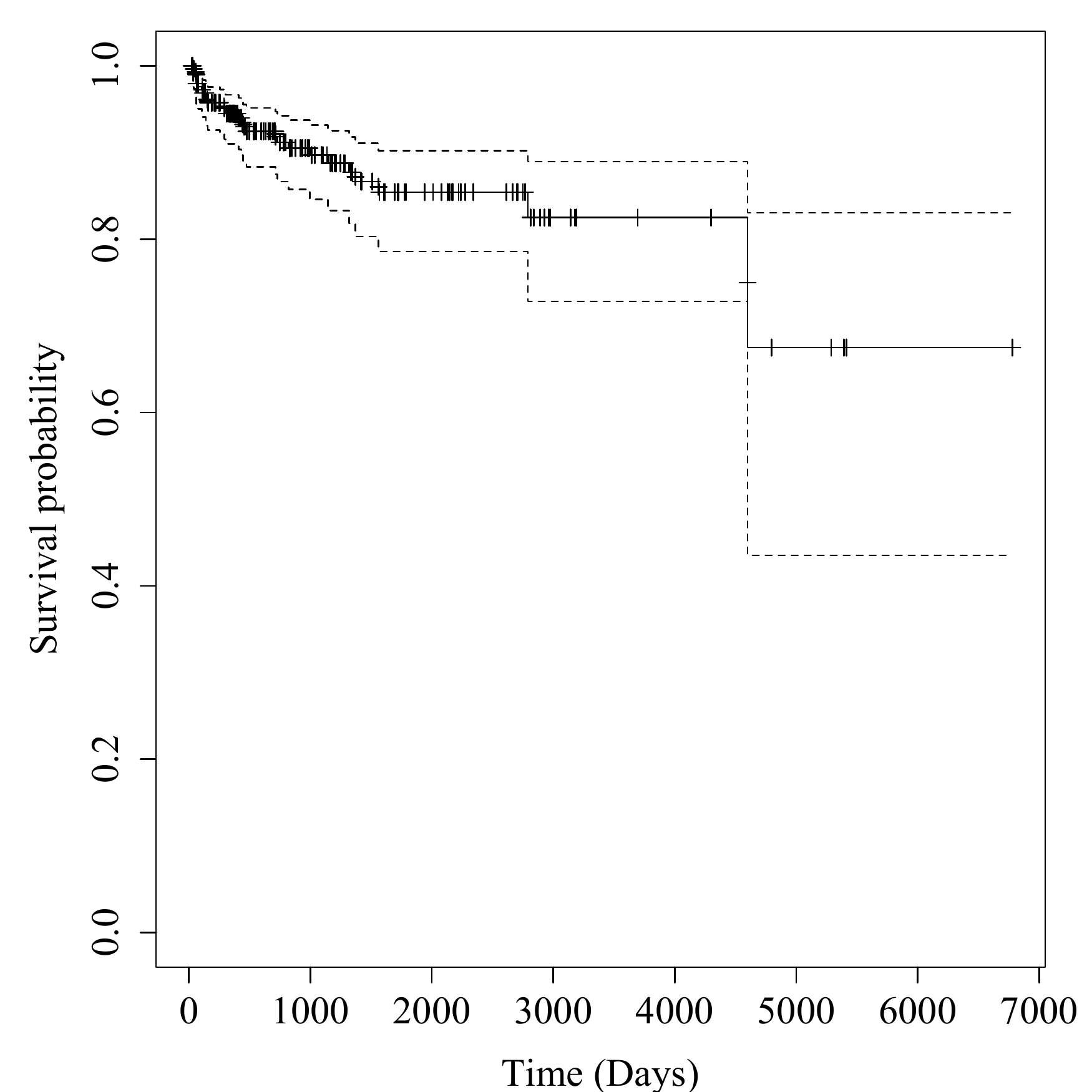}
    \caption{Kaplan-Meier survival curve for \textit{Implicit Columns} SQL code smell. The X-axis is the time in days and the vertical axis shows the survival probability value. The Censoring time and the Confidence interval are marked in the plot.}
    \label{fig:RQ4_imp_col}
    \vspace{-6pt}
\end{figure}

We perform survival analysis \cite{miller2011survival} to determine how long the SQL code smells survive throughout the life cycle of a subject system. Fig. \ref{fig:RQ4_imp_col} shows the Kaplan-Meier survival curve of \textit{Implicit Columns} SQL code smells against 150 data-intensive subject systems.  
We see that the survival curve has a steeper slope at the beginning and becomes flat after 3000 days. This indicates that a large fraction of this smell was either fixed or censored without getting fixed in this time. However, the large number of censored data points indicate that a significant part of \textit{Implicit Columns} smells persist without getting fixed.
%the event of interest (e.g., removal of SQL code smells, refactoring) have happened during these days. 
%Furthermore, all \textit{Implicit Columns} smells have survival probability greater than 0.6. 
%But equally proportional files took more longer days to get fixed. 
%When we analyse the distribution of survived days, we found that  the median survival time of \textit{Implicit Columns} smell is 1,501 days, which is approximately four years.
  
\begin{figure}[!ht]
    
    \includegraphics[width=2.4in]{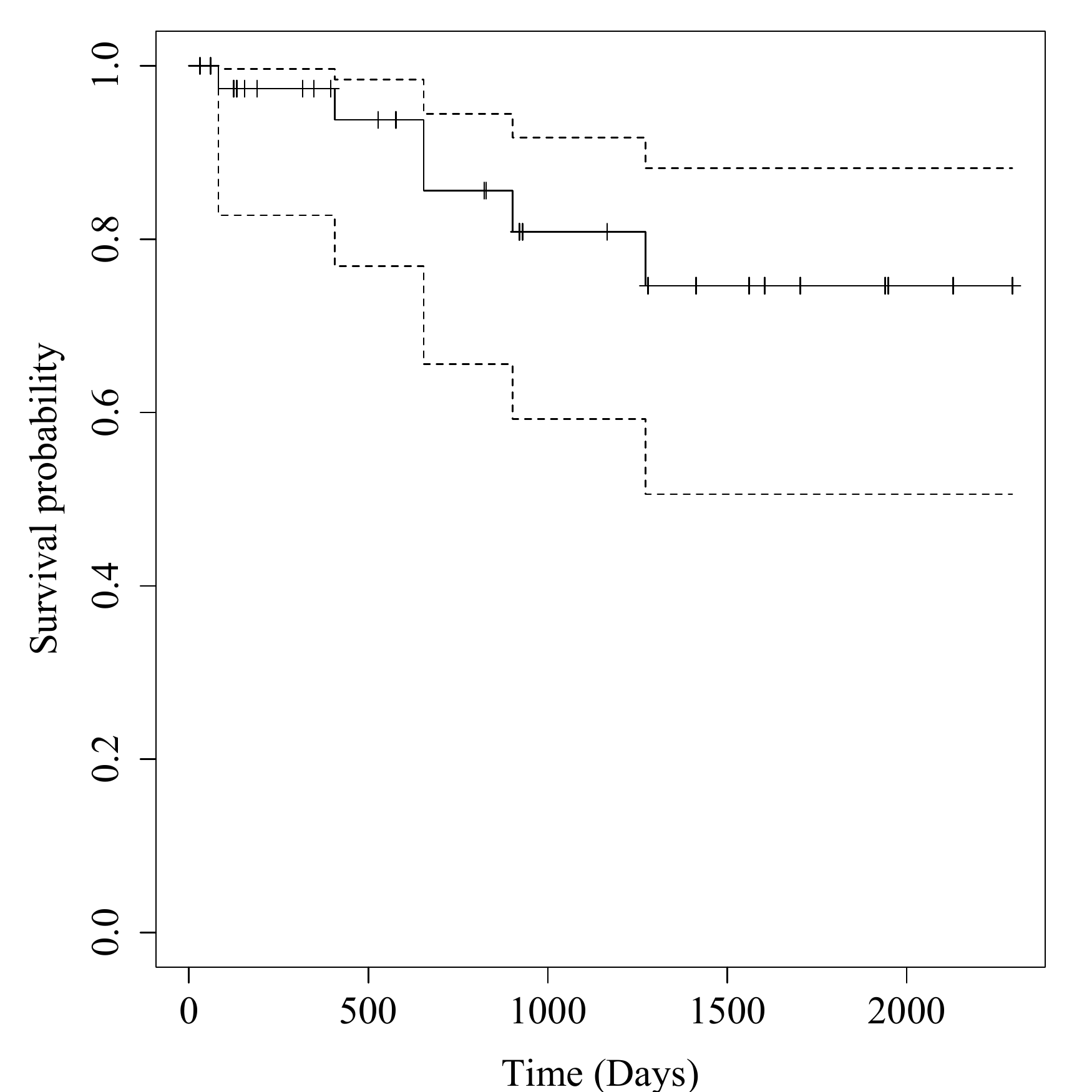}
    \caption{Kaplan-Meier survival curve for \textit{Fear of the Unknown} SQL code smell. The X-axis is the time in days and the vertical axis shows the survival probability value. The Censoring time and the Confidence interval are marked in the plot.}
    \label{fig:RQ4_fear_unknown}
    \vspace{-6pt}
\end{figure}

Fig. \ref{fig:RQ4_fear_unknown} shows the Kaplan-Meier
survival curve for another prevalent SQL code smell namely  \textit{Fear of the Unknown}. 
%We see that the shape of this curve is different from that of \textit{Implicit Columns}.
%as shown in Figure \ref{fig:RQ4_fear_unknown}. 
It has a similar trend to that of \textit{Implicit Columns} but the events are more visible due to small number of instances of this smell in the dataset.   

%at the beginning, which indicates that a small portion of these smells are fixed within a short time duration. However, a large fraction of the files affected with the smell took between 1,000 and 2,000 days to get fixed. 

%The median survival time of \textit{Fear of the Unknown} smells is 1,515 days (i.e., approximately 4 years) which is 14 days higher than that of \textit{Implicit Columns}.

\begin{figure}[!ht]
    \vspace{6pt}
    \includegraphics[width=2.4in]{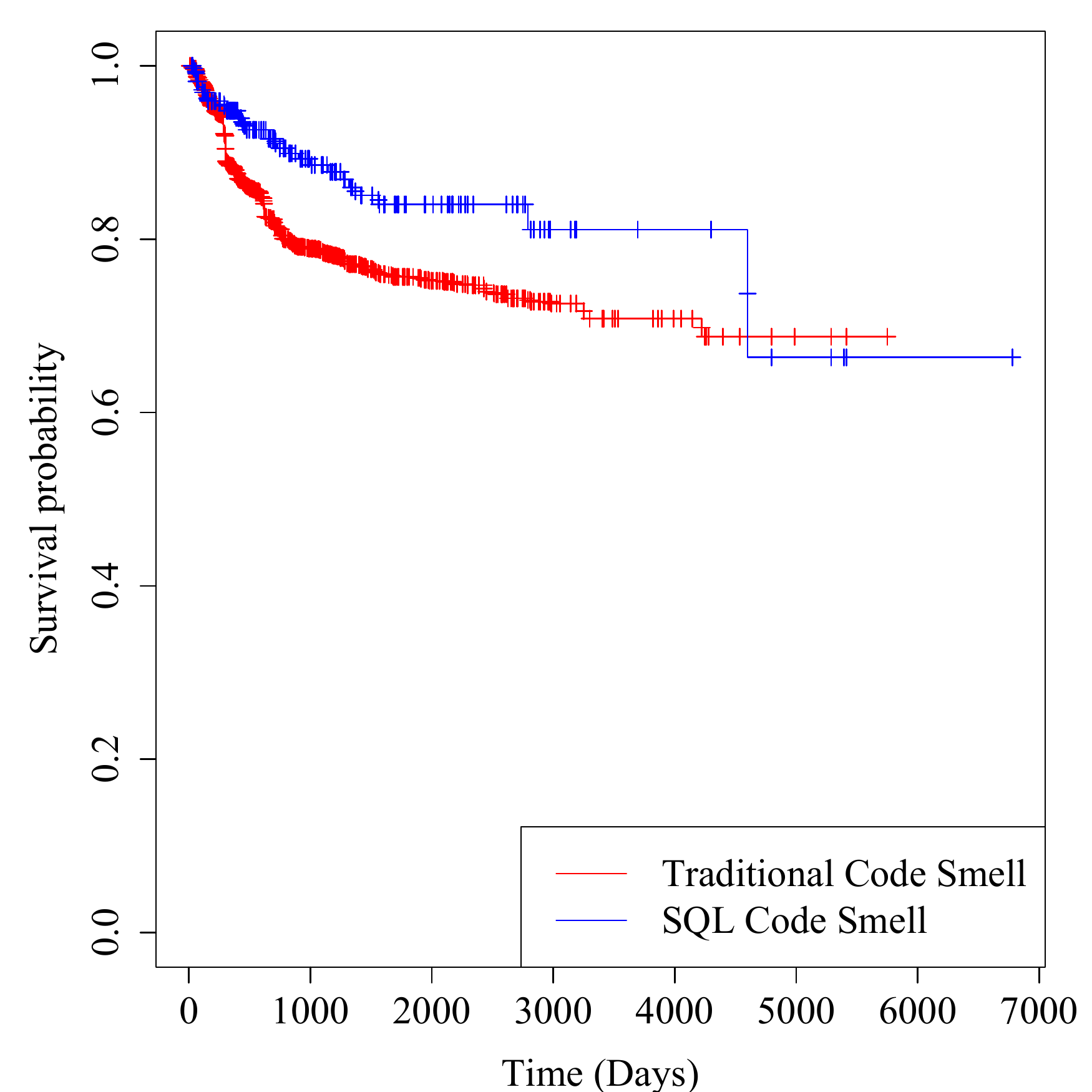}
    \caption{Kaplan-Meier survival curve for traditional code smells and SQL code smells. The Censoring time for censored files is marked in the plot.}
    \label{fig:RQ4_trad_vs_database}
    \vspace{-17pt}
\end{figure}

In order to achieve further insights, we compare the survival time of SQL code smells with that of traditional code smells. We run survival analysis on the two most prevalent traditional smells that are \textit{LongMethod} and \textit{LongParameterList}.
%It is interesting to compare the survival time of traditional code smells and database smells. 
Fig. \ref{fig:RQ4_trad_vs_database} shows our comparative analysis between 
traditional and SQL code smells. 
We see that SQL code smells have gentler survival curve than that of traditional smells. That is, SQL code smells have longer lifespan. Thus, they persist within the subject systems for a longer time duration.

%Based on our  distribution analysis on survival information, they have a median survival time of 1,501 days (four years). On the contrary, traditional code smells have a median survival time of 3,413 days, which is more than nine years. Thus, they persist within the subject systems for a longer time duration.
%(more than 9 years) which is more than double. 
We performed \emph{Logrank test} \cite{peto1972asymptotically} to determine whether the difference between these two survival curves in Fig. \ref{fig:RQ4_trad_vs_database} is statistically significant or not. We obtained a Chi-squared test p-value of $0.002$, which provides a strong evidence that these survival curves are significantly different. In both curves, we see large number of censored data. By censored we mean those files whose smells either persist in all tracked snapshots or they are deleted from the projects during the observation window which is a rare case in our data. 

%indicates that there is strong evidence that the difference in survival curves are statistically significant. 
%In both cases, there is  which indicates a large number of database smells as well as traditional code smells persist on all tracked versions without getting fixed.

We also track the SQL code smells that occur across the versions of each single subject systems. Based on our investigation, we found that a large percentage of SQL code smells occurred in early versions of the subject systems.
%oldest version 
%and the latest version of tracked projects. 
For instance, 89.5\% of the source code files with \textit{Implicit Columns} had their smells introduced in their first tracked snapshots. 
%\Foutse{what is oldest? is it the first release? i doesn't seem so since you use a different term for \textit{Fear of the Unknown}, please clarify} version. 
Similarly, 72.5\% of the source code files with \textit{Fear of the Unknown} had their smells introduced in their first tracked snapshot. 

Our analysis shows that, 80.5\% of source code files with \textit{Implicit Columns} smell contained this smell in all snapshots. Similarly, 65\% of source code files with the  \textit{Fear of the Unknown} smell contained this smell in all snapshots. In contrast,  54\% of source code files with \textit{LongParameterList} and 65\% of source code files with the \textit{LongMethod} contain those traditional code smells in all snapshots. This confirms the observation that a large number of files with SQL code smells and traditional code smells were censored before they are getting fixed. All these findings above suggest that SQL code smells get a little to no attention from the developers for refactoring.

%\Foutse{meaning?}
%This shows that database smells get smaller consideration for refactoring compared to traditional code smells. 

\vspace{0.2cm}
\begin{tcolorbox}[colback=white, colframe=black,left=2pt,right=2pt,top=1pt,bottom=1pt]
%The median survival time of SQL code smells is four years which is shorter compared to that of traditional code smells (e.g., nine years). However,
\textbf{SQL code smells have higher tendency to survive for longer period of time compared to traditional code smells.
A large fraction of the source files affected by SQL code smells (80.5\%) persist throughout the whole snapshots, and they hardly get any attention from the developers during refactoring.}
%stay for the whole versions without getting refactored compared to traditional code smells. 
\end{tcolorbox}

% -----------------------------------------------------------------------------------------------------------------------------------------------------------------
%\vspace{-10pt}
\vspace{-6pt}
\section{Implication of findings}
\label{sec:implication}

%\subsection{RQ1: prevalence} 
The result of RQ1 shows that not all SQL code smells are equally prevalent in data-intensive projects. Developers need to focus their attention on smells that are prevalent such as \textit{Implicit Columns} which
%The prevalence of \textit{Implicit Columns} smell shows that developers often use select queries without explicitly specifying column names to simplify the query. However, it 
may lead to unexpected issues in the production environment. The prevalence of SQL code smells on Library projects is more concerning as it may propagate to other application domains.

%\subsection{RQ2: Co-occurrence between database smells and traditional code smells}
Our findings show a small but statistically significant co-occurrence between some SQL code smells and some traditional code smells. This result can be a starting point for investigation of the relation between SQL code smells and traditional code smells and potentially detecting the occurrence of SQL code smells given some traditional code smells or vice versa.

%\subsection{RQ3: Co-occurrence with bugs}
We did not see a strong co-occurrence between SQL code smells and bugs. Our result shows that some traditional code smells have a higher association with bugs compared to SQL code smells. This implies that, future investigation should focus on the impact of SQL code smells on maintainability and performance instead of their link with bugs.

%\subsection{RQ4: Survival of database smells} 

The result of RQ4 shows that little attention is given to SQL code smells. Large portions of those smells are created in the first tracked snapshot of our subject systems and tend to persist for longer period of time. This implies that smells in general and SQL code smells in particular get a low priority in refactoring. The reasons for this could be developers' lack of awareness about those smells and their potential negative impact, or developers' engagement in higher priority tasks such as bug fixing tasks. 

%On average database smells survive for more than four years while traditional code smells may survive more than nine years. However, a high percentage of SQL code smells stayed in the whole version history of our projects, while it was not the case for  traditional code smells. This demonstrates that there is a very small attempt by developers to refactor both SQL code smells and traditional code smells.

\section{Threats to validity}
\label{sec:threats}
\vspace{6pt}
\textbf{\textbf{Threats to construct validity:}} We relied on the accuracy of SQLInspect and DECOR detection tools. Both tools may miss some smells. 
While the results reflect the minimum case, the actual number of smells could be higher. We used git diff for file history tracking, which might fail to
track some files 
%however, some files may not be tracked 
if they are moved using \textit{mv} command instead of \textit{git move}. We did not include such files in our study. We also used a 70\% similarity threshold for rename detection, which may lead to false rename assumption in some cases. However, the same threshold was used by the literature (e.g., by \citet{johannes2019large}). To link bugs with file versions, we relied on the SZZ algorithm, which might not be free from limitations. First, the heuristics of finding bug-fix commits using keywords may introduce false positives, which might incorrectly identify buggy lines \cite{rodriguez2018reproducibility}.  We also manually checked 50 randomly-sampled, bug-inducing commits detected by the SZZ algorithm and found only three (6\%) false-positives.
Thus, the threat posed by SZZ might not be significant. 
%We are forced to use such keywords as the commit style of each project is very different. However, we selected the keywords that are used in related studies to identify bug-fix commits.

\textbf{Threats to internal validity}: We did not claim any causation as this is an exploratory study. We only discussed the co-occurrence or association. Hence, our study is not subjected to threats to internal validity.

\textbf{Threats to conclusion validity}: To avoid conclusion threats to validity we only used non-parametric statistical tests.

\textbf{Threats to external validity}: To make our findings generalizable, we selected different types of projects in terms of application domain, size, and number of interactions with a database. We also covered projects that use different drivers and frameworks to interact with the database. We also tried to select representative projects with relevant data access. We only considered Java projects for analysis. However, our investigation approach is generalizable to any programming language. It is desirable to study if our conclusions can be extended to different programming languages.

\textbf{Threats to reliability validity}: To minimize potential threats to reliability, we analyzed open-source projects available on GitHub and provide a replication package that contains our dataset \cite{replication}.

% -----------------------------------------------------------------------------------------------------------------------------------------------------------------

\section{Related work}
\label{sec:related}
We discuss the state of the art from two perspectives. First, we overview the related work on empirical studies about traditional code smells. Then, we discuss the state of the art research on SQL code or database-related smells.

\textbf{Traditional Code Smells:} Since the initial introduction of the term ``design flaws'' \cite{Riel1996} and ``code smell'' \cite{FowlerRefactoring}, many studied their impact on development; i.e., how they affect performance, source code quality or maintainability. A recent literature review ``on the code smell effect'' by Santos et al. \cite{Santos2018} gives an overview of these studies. They identify a total number of 3530 papers in this area and after removing duplicated and short papers they do an in-depth examination of 64 papers in their survey. Important to mention here are \textit{correlation studies} on code smells and quality attribute such as number of bugs or number of modifications in classes \cite{Yamashita2013,Fontana2013, Macia2012,zazworka2011investigating,li2007empirical} and empirical studies on the evolution of code smells \cite{peters2012evaluating,johannes2019large,olbrich2009evolution,khomh2009exploratory,shatnawi2006investigation,palomba2018diffuseness}, their influence on defects \cite{Olbrich2010}, maintenance effort \cite{Sjoberg2013} or how humans perceive them \cite{Mantyla2006, Palomba2014}. 

For the detection of traditional code smells, we rely on the DECOR tool \cite{gueheneuc2007ptidej}. However, different traditional smell detection tools are also available. For example, an artificial immune system-based smell detection tool was developed by Hassaine et al. \cite{hassaine2010ids}, a Bayesian network-based expert system by Khomh et al. \cite{khomh2011bdtex}, and a textual data mining approach by Palomba et al. \cite{palomba2015textual}. Kessentini and Ouni proposed an approach to automatically generate smell detection rules using genetic algorithm \cite{kessentini2017detecting}. Another popular traditional smell detection tool is JDeodorant developed by Tsantalis \cite{tsantalis2010evaluation}. JDeodorant focuses on the detection and refactoring of Feature Envy, God Class and Duplicated Code, Type Checking and Long Method smells. We choose DECOR because it covers more traditional code smells and was reported to reach a 100\% recall.

%The evolution of traditional code smells was studied by authors such as of Peters et.al \cite{peters2012evaluating}, Jonannes et.al \cite{johannes2019large} and Olbrich et al. \cite{olbrich2009evolution}. The impact of traditional code smells on software quality was addressed in some studies (Eg.  \cite{khomh2009exploratory,shatnawi2006investigation,palomba2018diffuseness,johannes2019large}).

\textbf{SQL Code Smells:} Although researchers studied common errors in SQL queries before \cite{Brass2004, Goldberg2009}, the book of Karwin \cite{karwin2010sql} is the first to present SQL antipatterns in a comprehensive catalogue. This catalogue inspired researchers to further investigate such smells. Khumnin et al. \cite{Khumnin2017} present a tool for detecting logical database design antipatterns in Transact-SQL queries. Nagy and Cleve \cite{nagy2017static} propose a static analysis approach to detect SQL code smells in queries extracted from Java code. They also provide additional analyses (e.g., metrics) about the detected smells \cite{nagy2018sqlinspect}.
Another tool, DbDeo \cite{Sharma2018}, implements the detection of \emph{database schema} smells. DbDeo is evaluated on 2925 open-source repositories; their authors identified 13 different types of smells, among which `index abuse' was found to be the most prevalent one. In another recent work, De Almeida Filho et al. \cite{de2019prevalence} investigate the prevalence and co-occurrence of SQL code smells in PL/SQL projects. Arzamasova et al. propose to detect antipatterns in SQL logs \cite{Arzamasova2018} and demonstrate their approach through the refactoring of a project containing more than 40 millions of queries. Let us also mention the work by Burzanska et al. \cite{Burzanska2018}, who question whether the `Poor Man's Search Engine' smell should still be considered as a poor practice today, as relational databases evolved since Karwin's catalogue. % Moreover, although this smell is one of the SQL Query smells in the catalogue, developers of SQLInspect opted not to implement its detection.

There exist several other SQL analysis and smell detection tools, including TOAD\footnote{\url{http://www.toadworld.com/}}, SQL Prompt\footnote{\url{https://www.red-gate.com/hub/product-learning/sql-prompt}}, and SQL Enlight\footnote{\url{https://sqlenlight.com/}}. However, those tools require a set of SQL queries as input, and they cannot analyze the queries embedded in source code. SQLInspect can extract the queries and detect SQL code smells given the project source code, which justifies our choice.

\textbf{Summary:} In contrast with the studies discussed above, this constitutes -- to the best of our knowledge -- the first empirical study investigating the prevalence of SQL code smells, their association with bugs and with other traditional code smells, as well as their evolution over time. We expect this study to serve as a baseline for further studies on the impact and persistence of SQL code smells in data-intensive systems.

\vspace{6pt}
\section{Conclusion and future work}
\label{sec:conclusion}

In this study, we investigated the prevalence of SQL code smells and their association with bugs and other traditional code smells. 
We collected 150 open-source Java projects, extracted both SQL and traditional code smells, and then jointly analyzed their prevalence and co-occurrence. We linked bug-inducing commits to those smells using the SZZ algorithm to study their association with bugs. We performed a survival analysis to study how SQL code smells are handled throughout the lifetime of these projects.

%We identified Co-occurring database smells and traditional code smells using Apriori algorithm. We also conducted statistical test of association to study co-occurrence of database smells with traditional code smells and with bugs. We modeled the survival time of database smells and traditional code smells using Kaplan-Meier survival function.

Our results show that SQL code smells are prevalent in open-source data-intensive systems, but at different levels. In particular, we found that the \textit{Implicit Columns} SQL code smell is the most prevalent in our subject systems. With some exceptions, however, we did not see a significant difference in the prevalence of SQL code smells among application domains. Also, we found only a weak association between SQL code smells and traditional code smells. 
%Database smells have smaller association with bugs compared to traditional code smells. 
Survival analysis showed that 
%SQL code smells have a median survival time of more than four years while traditional code smells have median survival time of nine years. However, 
a significant portion of SQL code smells was created in the first tracked snapshot of the studied systems and persisted in all snapshots without getting fixed.

Overall, our findings indicate that SQL code smells exist persistently in data-intensive systems, but independently from traditional code smells. As a consequence, developers have to be aware of SQL code smells, so that they can identify those smells and refactor them in order to avoid potential harm.
%Based on our results, we recommend developers to be aware of SQL code smells and consider refactoring SQL code smells for data-intensive applications to obtain optimal interaction with database systems. 

Our study is exploratory in nature. We believe that further investigation is needed to better understand the consequences of SQL code smells. This includes, in particular, their impact on the performance and the maintainability of data-intensive systems. %We also see the need for qualitative analyses of developers awareness about SQL code smells.

%% just to not forget
\textbf{Acknowledgements:} This research was partly supported by the Excellence of Science project 30446992 SECO-ASSIST, funded by the F.R.S.-FNRS, FWO and Natural Sciences and Engineering Research Council of Canada (NSERC).

%\newpage
\bibliographystyle{ACM-Reference-Format}
\bibliography{Citations}

%%% -*-BibTeX-*-
%%% Do NOT edit. File created by BibTeX with style
%%% ACM-Reference-Format-Journals [18-Jan-2012].

\begin{thebibliography}{63}

%%% ====================================================================
%%% NOTE TO THE USER: you can override these defaults by providing
%%% customized versions of any of these macros before the \bibliography
%%% command.  Each of them MUST provide its own final punctuation,
%%% except for \shownote{}, \showDOI{}, and \showURL{}.  The latter two
%%% do not use final punctuation, in order to avoid confusing it with
%%% the Web address.
%%%
%%% To suppress output of a particular field, define its macro to expand
%%% to an empty string, or better, \unskip, like this:
%%%
%%% \newcommand{\showDOI}[1]{\unskip}   % LaTeX syntax
%%%
%%% \def \showDOI #1{\unskip}           % plain TeX syntax
%%%
%%% ====================================================================

\ifx \showCODEN    \undefined \def \showCODEN     #1{\unskip}     \fi
\ifx \showDOI      \undefined \def \showDOI       #1{#1}\fi
\ifx \showISBNx    \undefined \def \showISBNx     #1{\unskip}     \fi
\ifx \showISBNxiii \undefined \def \showISBNxiii  #1{\unskip}     \fi
\ifx \showISSN     \undefined \def \showISSN      #1{\unskip}     \fi
\ifx \showLCCN     \undefined \def \showLCCN      #1{\unskip}     \fi
\ifx \shownote     \undefined \def \shownote      #1{#1}          \fi
\ifx \showarticletitle \undefined \def \showarticletitle #1{#1}   \fi
\ifx \showURL      \undefined \def \showURL       {\relax}        \fi
% The following commands are used for tagged output and should be
% invisible to TeX
\providecommand\bibfield[2]{#2}
\providecommand\bibinfo[2]{#2}
\providecommand\natexlab[1]{#1}
\providecommand\showeprint[2][]{arXiv:#2}

\bibitem[\protect\citeauthoryear{Agarwal}{Agarwal}{2017}]%
        {agarwal2017decision}
\bibfield{author}{\bibinfo{person}{R Agarwal}.}
  \bibinfo{year}{2017}\natexlab{}.
\newblock \showarticletitle{Decision making with association rule mining and
  clustering in supply chains}.
\newblock \bibinfo{journal}{\emph{International Journal of Data and Network
  Science}} \bibinfo{volume}{1}, \bibinfo{number}{1} (\bibinfo{year}{2017}),
  \bibinfo{pages}{11--18}.
\newblock


\bibitem[\protect\citeauthoryear{Agrawal, Imielinski, and Swami}{Agrawal
  et~al\mbox{.}}{1993}]%
        {agrawal1993mining}
\bibfield{author}{\bibinfo{person}{R Agrawal}, \bibinfo{person}{T Imielinski},
  {and} \bibinfo{person}{A Swami}.} \bibinfo{year}{1993}\natexlab{}.
\newblock \showarticletitle{Mining associations between sets of items in large
  databases}. In \bibinfo{booktitle}{\emph{Proceedings of the ACM SIGMOD
  International Conference on Management of Data}}. \bibinfo{pages}{207--216}.
\newblock


\bibitem[\protect\citeauthoryear{Agrawal, Srikant, et~al\mbox{.}}{Agrawal
  et~al\mbox{.}}{1994}]%
        {agrawal1994fast}
\bibfield{author}{\bibinfo{person}{Rakesh Agrawal},
  \bibinfo{person}{Ramakrishnan Srikant}, {et~al\mbox{.}}}
  \bibinfo{year}{1994}\natexlab{}.
\newblock \showarticletitle{Fast algorithms for mining association rules}. In
  \bibinfo{booktitle}{\emph{Proc. 20th Int. Conf. Very Large Data Bases,
  VLDB}}, Vol.~\bibinfo{volume}{1215}. \bibinfo{pages}{487--499}.
\newblock


\bibitem[\protect\citeauthoryear{AlZu'bi, Hawashin, EIBes, and
  Al-Ayyoub}{AlZu'bi et~al\mbox{.}}{2018}]%
        {alzu2018novel}
\bibfield{author}{\bibinfo{person}{Shadi AlZu'bi}, \bibinfo{person}{Bilal
  Hawashin}, \bibinfo{person}{Mohammad EIBes}, {and} \bibinfo{person}{Mahmoud
  Al-Ayyoub}.} \bibinfo{year}{2018}\natexlab{}.
\newblock \showarticletitle{A novel recommender system based on apriori
  algorithm for requirements engineering}. In \bibinfo{booktitle}{\emph{2018
  Fifth International Conference on Social Networks Analysis, Management and
  Security (SNAMS)}}. IEEE, \bibinfo{pages}{323--327}.
\newblock


\bibitem[\protect\citeauthoryear{Aniche, Bavota, Treude, Gerosa, and van
  Deursen}{Aniche et~al\mbox{.}}{2018}]%
        {aniche2018code}
\bibfield{author}{\bibinfo{person}{Maur{\'\i}cio Aniche},
  \bibinfo{person}{Gabriele Bavota}, \bibinfo{person}{Christoph Treude},
  \bibinfo{person}{Marco~Aur{\'e}lio Gerosa}, {and} \bibinfo{person}{Arie van
  Deursen}.} \bibinfo{year}{2018}\natexlab{}.
\newblock \showarticletitle{Code smells for model-view-controller
  architectures}.
\newblock \bibinfo{journal}{\emph{Empirical Software Engineering}}
  \bibinfo{volume}{23}, \bibinfo{number}{4} (\bibinfo{year}{2018}),
  \bibinfo{pages}{2121--2157}.
\newblock


\bibitem[\protect\citeauthoryear{Antoniol, Ayari, Di~Penta, Khomh, and
  Gu{\'e}h{\'e}neuc}{Antoniol et~al\mbox{.}}{2008}]%
        {antoniol2008bug}
\bibfield{author}{\bibinfo{person}{Giuliano Antoniol}, \bibinfo{person}{Kamel
  Ayari}, \bibinfo{person}{Massimiliano Di~Penta}, \bibinfo{person}{Foutse
  Khomh}, {and} \bibinfo{person}{Yann-Ga{\"e}l Gu{\'e}h{\'e}neuc}.}
  \bibinfo{year}{2008}\natexlab{}.
\newblock \showarticletitle{Is it a bug or an enhancement? A text-based
  approach to classify change requests}. In \bibinfo{booktitle}{\emph{CASCON}},
  Vol.~\bibinfo{volume}{8}. \bibinfo{pages}{304--318}.
\newblock


\bibitem[\protect\citeauthoryear{{Arzamasova}, {Schäler}, and
  {Böhm}}{{Arzamasova} et~al\mbox{.}}{2018}]%
        {Arzamasova2018}
\bibfield{author}{\bibinfo{person}{N. {Arzamasova}}, \bibinfo{person}{M.
  {Schäler}}, {and} \bibinfo{person}{K. {Böhm}}.}
  \bibinfo{year}{2018}\natexlab{}.
\newblock \showarticletitle{Cleaning Antipatterns in an SQL Query Log}.
\newblock \bibinfo{journal}{\emph{IEEE Transactions on Knowledge and Data
  Engineering}} \bibinfo{volume}{30}, \bibinfo{number}{3}
  (\bibinfo{date}{March} \bibinfo{year}{2018}), \bibinfo{pages}{421--434}.
\newblock
\showISSN{2326-3865}


\bibitem[\protect\citeauthoryear{{Brass} and {Goldberg}}{{Brass} and
  {Goldberg}}{2004}]%
        {Brass2004}
\bibfield{author}{\bibinfo{person}{S. {Brass}} {and} \bibinfo{person}{C.
  {Goldberg}}.} \bibinfo{year}{2004}\natexlab{}.
\newblock \showarticletitle{Semantic errors in SQL queries: a quite complete
  list}. In \bibinfo{booktitle}{\emph{Fourth International Conference on
  Quality Software, 2004. QSIC 2004. Proceedings.}} \bibinfo{pages}{250--257}.
\newblock


\bibitem[\protect\citeauthoryear{Brin, Motwani, Ullman, and Tsur}{Brin
  et~al\mbox{.}}{1997}]%
        {brin1997dynamic}
\bibfield{author}{\bibinfo{person}{Sergey Brin}, \bibinfo{person}{Rajeev
  Motwani}, \bibinfo{person}{Jeffrey~D Ullman}, {and} \bibinfo{person}{Shalom
  Tsur}.} \bibinfo{year}{1997}\natexlab{}.
\newblock \showarticletitle{Dynamic itemset counting and implication rules for
  market basket data}.
\newblock \bibinfo{journal}{\emph{Acm Sigmod Record}} \bibinfo{volume}{26},
  \bibinfo{number}{2} (\bibinfo{year}{1997}), \bibinfo{pages}{255--264}.
\newblock


\bibitem[\protect\citeauthoryear{Burza{\'{n}}ska and
  Wi{\'{s}}niewski}{Burza{\'{n}}ska and Wi{\'{s}}niewski}{2018}]%
        {Burzanska2018}
\bibfield{author}{\bibinfo{person}{Marta Burza{\'{n}}ska} {and}
  \bibinfo{person}{Piotr Wi{\'{s}}niewski}.} \bibinfo{year}{2018}\natexlab{}.
\newblock \showarticletitle{How Poor Is the ``Poor Man's Search Engine''?}. In
  \bibinfo{booktitle}{\emph{Beyond Databases, Architectures and Structures.
  Facing the Challenges of Data Proliferation and Growing Variety}},
  \bibfield{editor}{\bibinfo{person}{Stanis{\l}aw Kozielski},
  \bibinfo{person}{Dariusz Mrozek}, \bibinfo{person}{Pawe{\l} Kasprowski},
  \bibinfo{person}{Bo{\.{z}}ena Ma{\l}ysiak-Mrozek}, {and}
  \bibinfo{person}{Daniel Kostrzewa}} (Eds.). \bibinfo{publisher}{Springer
  International Publishing}, \bibinfo{address}{Cham},
  \bibinfo{pages}{294--305}.
\newblock
\showISBNx{978-3-319-99987-6}


\bibitem[\protect\citeauthoryear{Chawla, Bowyer, Hall, and Kegelmeyer}{Chawla
  et~al\mbox{.}}{2002}]%
        {chawla2002smote}
\bibfield{author}{\bibinfo{person}{Nitesh~V Chawla}, \bibinfo{person}{Kevin~W
  Bowyer}, \bibinfo{person}{Lawrence~O Hall}, {and} \bibinfo{person}{W~Philip
  Kegelmeyer}.} \bibinfo{year}{2002}\natexlab{}.
\newblock \showarticletitle{SMOTE: synthetic minority over-sampling technique}.
\newblock \bibinfo{journal}{\emph{Journal of Artificial Intelligence Research}}
   \bibinfo{volume}{16} (\bibinfo{year}{2002}), \bibinfo{pages}{321--357}.
\newblock


\bibitem[\protect\citeauthoryear{Cramer}{Cramer}{1946}]%
        {cramir1946mathematical}
\bibfield{author}{\bibinfo{person}{Harald Cramer}.}
  \bibinfo{year}{1946}\natexlab{}.
\newblock \showarticletitle{Mathematical methods of statistics}.
\newblock \bibinfo{journal}{\emph{Princeton U. Press, Princeton}}
  (\bibinfo{year}{1946}), \bibinfo{pages}{500}.
\newblock


\bibitem[\protect\citeauthoryear{de~Almeida~Filho, Martins, Vinuto, Monteiro,
  de~Sousa, de~Castro~Machado, and Rocha}{de~Almeida~Filho
  et~al\mbox{.}}{2019}]%
        {de2019prevalence}
\bibfield{author}{\bibinfo{person}{Francisco~Gon{\c{c}}alves de Almeida~Filho},
  \bibinfo{person}{Ant{\^o}nio Diogo~Forte Martins}, \bibinfo{person}{Tiago
  da~Silva Vinuto}, \bibinfo{person}{Jos{\'e}~Maria Monteiro},
  \bibinfo{person}{{\'I}talo~Pereira de Sousa}, \bibinfo{person}{Javam de
  Castro~Machado}, {and} \bibinfo{person}{Lincoln~Souza Rocha}.}
  \bibinfo{year}{2019}\natexlab{}.
\newblock \showarticletitle{Prevalence of bad smells in PL/SQL projects}. In
  \bibinfo{booktitle}{\emph{Proceedings of the 27th International Conference on
  Program Comprehension}}. IEEE Press, \bibinfo{pages}{116--121}.
\newblock


\bibitem[\protect\citeauthoryear{{Fontana}, {Ferme}, {Marino}, {Walter}, and
  {Martenka}}{{Fontana} et~al\mbox{.}}{2013}]%
        {Fontana2013}
\bibfield{author}{\bibinfo{person}{F.~A. {Fontana}}, \bibinfo{person}{V.
  {Ferme}}, \bibinfo{person}{A. {Marino}}, \bibinfo{person}{B. {Walter}}, {and}
  \bibinfo{person}{P. {Martenka}}.} \bibinfo{year}{2013}\natexlab{}.
\newblock \showarticletitle{Investigating the Impact of Code Smells on System's
  Quality: An Empirical Study on Systems of Different Application Domains}. In
  \bibinfo{booktitle}{\emph{2013 IEEE International Conference on Software
  Maintenance}}. \bibinfo{pages}{260--269}.
\newblock
\showISSN{1063-6773}


\bibitem[\protect\citeauthoryear{Fowler, Beck, Brant, Opdyke, Roberts, and
  Gamma}{Fowler et~al\mbox{.}}{1999}]%
        {FowlerRefactoring}
\bibfield{author}{\bibinfo{person}{Martin Fowler}, \bibinfo{person}{Kent Beck},
  \bibinfo{person}{John Brant}, \bibinfo{person}{William Opdyke},
  \bibinfo{person}{Don Roberts}, {and} \bibinfo{person}{Erich Gamma}.}
  \bibinfo{year}{1999}\natexlab{}.
\newblock \bibinfo{booktitle}{\emph{Refactoring: Improving the Design of
  Existing Code}}.
\newblock \bibinfo{publisher}{Addison-Wesley Longman Publishing Co., Inc.},
  \bibinfo{address}{USA}.
\newblock
\showISBNx{0201485672}


\bibitem[\protect\citeauthoryear{Goldberg}{Goldberg}{2009}]%
        {Goldberg2009}
\bibfield{author}{\bibinfo{person}{Christian Goldberg}.}
  \bibinfo{year}{2009}\natexlab{}.
\newblock \showarticletitle{Do You Know SQL? About Semantic Errors in Database
  Queries}. In \bibinfo{booktitle}{\emph{In 7th Workshop on Teaching, Learning
  and Assessment in Databases}}. \bibinfo{pages}{13--19}.
\newblock


\bibitem[\protect\citeauthoryear{Gu{\'e}h{\'e}neuc}{Gu{\'e}h{\'e}neuc}{2007}]%
        {gueheneuc2007ptidej}
\bibfield{author}{\bibinfo{person}{Yann-Ga{\"e}l Gu{\'e}h{\'e}neuc}.}
  \bibinfo{year}{2007}\natexlab{}.
\newblock \showarticletitle{Ptidej: A flexible reverse engineering tool suite}.
  In \bibinfo{booktitle}{\emph{2007 IEEE International Conference on Software
  Maintenance}}. IEEE, \bibinfo{pages}{529--530}.
\newblock


\bibitem[\protect\citeauthoryear{Guerrouj, Kermansaravi, Arnaoudova, Fung,
  Khomh, Antoniol, and Gu{\'e}h{\'e}neuc}{Guerrouj et~al\mbox{.}}{2017}]%
        {guerrouj2017investigating}
\bibfield{author}{\bibinfo{person}{Latifa Guerrouj}, \bibinfo{person}{Zeinab
  Kermansaravi}, \bibinfo{person}{Venera Arnaoudova},
  \bibinfo{person}{Benjamin~CM Fung}, \bibinfo{person}{Foutse Khomh},
  \bibinfo{person}{Giuliano Antoniol}, {and} \bibinfo{person}{Yann-Ga{\"e}l
  Gu{\'e}h{\'e}neuc}.} \bibinfo{year}{2017}\natexlab{}.
\newblock \showarticletitle{Investigating the relation between lexical smells
  and change-and fault-proneness: an empirical study}.
\newblock \bibinfo{journal}{\emph{Software Quality Journal}}
  \bibinfo{volume}{25}, \bibinfo{number}{3} (\bibinfo{year}{2017}),
  \bibinfo{pages}{641--670}.
\newblock


\bibitem[\protect\citeauthoryear{Hassaine, Khomh, Gu{\'e}h{\'e}neuc, and
  Hamel}{Hassaine et~al\mbox{.}}{2010}]%
        {hassaine2010ids}
\bibfield{author}{\bibinfo{person}{Salima Hassaine}, \bibinfo{person}{Foutse
  Khomh}, \bibinfo{person}{Yann-Ga{\"e}l Gu{\'e}h{\'e}neuc}, {and}
  \bibinfo{person}{Sylvie Hamel}.} \bibinfo{year}{2010}\natexlab{}.
\newblock \showarticletitle{IDS: An immune-inspired approach for the detection
  of software design smells}. In \bibinfo{booktitle}{\emph{2010 Seventh
  International Conference on the Quality of Information and Communications
  Technology}}. IEEE, \bibinfo{pages}{343--348}.
\newblock


\bibitem[\protect\citeauthoryear{Hecht, Moha, and Rouvoy}{Hecht
  et~al\mbox{.}}{2016}]%
        {hecht2016empirical}
\bibfield{author}{\bibinfo{person}{Geoffrey Hecht}, \bibinfo{person}{Naouel
  Moha}, {and} \bibinfo{person}{Romain Rouvoy}.}
  \bibinfo{year}{2016}\natexlab{}.
\newblock \showarticletitle{An empirical study of the performance impacts of
  android code smells}. In \bibinfo{booktitle}{\emph{Proceedings of the
  International Conference on Mobile Software Engineering and Systems}}. ACM,
  \bibinfo{pages}{59--69}.
\newblock


\bibitem[\protect\citeauthoryear{Inc}{Inc}{2019}]%
        {github_code_search}
\bibfield{author}{\bibinfo{person}{GitHub Inc}.}
  \bibinfo{year}{2019}\natexlab{}.
\newblock \bibinfo{title}{Search}.
\newblock
\newblock
\urldef\tempurl%
\url{https://developer.github.com/v3/search/}
\showURL{%
Retrieved December 28, 2019 from \tempurl}


\bibitem[\protect\citeauthoryear{Jin, Cui, and Yan}{Jin et~al\mbox{.}}{2019}]%
        {jin2019survey}
\bibfield{author}{\bibinfo{person}{Zichuan Jin}, \bibinfo{person}{Yanpeng Cui},
  {and} \bibinfo{person}{Zheng Yan}.} \bibinfo{year}{2019}\natexlab{}.
\newblock \showarticletitle{Survey of Intrusion Detection Methods Based on Data
  Mining Algorithms}. In \bibinfo{booktitle}{\emph{Proceedings of the 2019
  International Conference on Big Data Engineering}}. ACM,
  \bibinfo{pages}{98--106}.
\newblock


\bibitem[\protect\citeauthoryear{Johannes, Khomh, and Antoniol}{Johannes
  et~al\mbox{.}}{2019}]%
        {johannes2019large}
\bibfield{author}{\bibinfo{person}{David Johannes}, \bibinfo{person}{Foutse
  Khomh}, {and} \bibinfo{person}{Giuliano Antoniol}.}
  \bibinfo{year}{2019}\natexlab{}.
\newblock \showarticletitle{A large-scale empirical study of code smells in
  JavaScript projects}.
\newblock \bibinfo{journal}{\emph{Software Quality Journal}}
  (\bibinfo{year}{2019}), \bibinfo{pages}{1--44}.
\newblock


\bibitem[\protect\citeauthoryear{Kamei, Shihab, Adams, Hassan, Mockus, Sinha,
  and Ubayashi}{Kamei et~al\mbox{.}}{2012}]%
        {kamei2012large}
\bibfield{author}{\bibinfo{person}{Yasutaka Kamei}, \bibinfo{person}{Emad
  Shihab}, \bibinfo{person}{Bram Adams}, \bibinfo{person}{Ahmed~E Hassan},
  \bibinfo{person}{Audris Mockus}, \bibinfo{person}{Anand Sinha}, {and}
  \bibinfo{person}{Naoyasu Ubayashi}.} \bibinfo{year}{2012}\natexlab{}.
\newblock \showarticletitle{A large-scale empirical study of just-in-time
  quality assurance}.
\newblock \bibinfo{journal}{\emph{IEEE Transactions on Software Engineering}}
  \bibinfo{volume}{39}, \bibinfo{number}{6} (\bibinfo{year}{2012}),
  \bibinfo{pages}{757--773}.
\newblock


\bibitem[\protect\citeauthoryear{Kaplan and Meier}{Kaplan and Meier}{1958}]%
        {kaplan1958nonparametric}
\bibfield{author}{\bibinfo{person}{Edward~L Kaplan} {and} \bibinfo{person}{Paul
  Meier}.} \bibinfo{year}{1958}\natexlab{}.
\newblock \showarticletitle{Nonparametric estimation from incomplete
  observations}.
\newblock \bibinfo{journal}{\emph{Journal of the American statistical
  association}} \bibinfo{volume}{53}, \bibinfo{number}{282}
  (\bibinfo{year}{1958}), \bibinfo{pages}{457--481}.
\newblock


\bibitem[\protect\citeauthoryear{Karwin}{Karwin}{2010}]%
        {karwin2010sql}
\bibfield{author}{\bibinfo{person}{Bill Karwin}.}
  \bibinfo{year}{2010}\natexlab{}.
\newblock \bibinfo{booktitle}{\emph{SQL Antipatterns: Avoiding the pitfalls of
  database programming}}.
\newblock \bibinfo{publisher}{Pragmatic Bookshelf}.
\newblock


\bibitem[\protect\citeauthoryear{Kaur and Kang}{Kaur and Kang}{2016}]%
        {kaur2016market}
\bibfield{author}{\bibinfo{person}{Manpreet Kaur} {and}
  \bibinfo{person}{Shivani Kang}.} \bibinfo{year}{2016}\natexlab{}.
\newblock \showarticletitle{Market Basket Analysis: Identify the changing
  trends of market data using association rule mining}.
\newblock \bibinfo{journal}{\emph{Procedia computer science}}
  \bibinfo{volume}{85} (\bibinfo{year}{2016}), \bibinfo{pages}{78--85}.
\newblock


\bibitem[\protect\citeauthoryear{Kessentini and Ouni}{Kessentini and
  Ouni}{2017}]%
        {kessentini2017detecting}
\bibfield{author}{\bibinfo{person}{Marouane Kessentini} {and}
  \bibinfo{person}{Ali Ouni}.} \bibinfo{year}{2017}\natexlab{}.
\newblock \showarticletitle{Detecting android smells using multi-objective
  genetic programming}. In \bibinfo{booktitle}{\emph{Proceedings of the 4th
  International Conference on Mobile Software Engineering and Systems}}. IEEE
  Press, \bibinfo{pages}{122--132}.
\newblock


\bibitem[\protect\citeauthoryear{Khomh, Di~Penta, and Gueheneuc}{Khomh
  et~al\mbox{.}}{2009}]%
        {khomh2009exploratory}
\bibfield{author}{\bibinfo{person}{Foutse Khomh}, \bibinfo{person}{Massimiliano
  Di~Penta}, {and} \bibinfo{person}{Yann-Gael Gueheneuc}.}
  \bibinfo{year}{2009}\natexlab{}.
\newblock \showarticletitle{An exploratory study of the impact of code smells
  on software change-proneness}. In \bibinfo{booktitle}{\emph{2009 16th Working
  Conference on Reverse Engineering}}. IEEE, \bibinfo{pages}{75--84}.
\newblock


\bibitem[\protect\citeauthoryear{Khomh, Vaucher, Gu{\'e}h{\'e}neuc, and
  Sahraoui}{Khomh et~al\mbox{.}}{2011}]%
        {khomh2011bdtex}
\bibfield{author}{\bibinfo{person}{Foutse Khomh}, \bibinfo{person}{Stephane
  Vaucher}, \bibinfo{person}{Yann-Ga{\"e}l Gu{\'e}h{\'e}neuc}, {and}
  \bibinfo{person}{Houari Sahraoui}.} \bibinfo{year}{2011}\natexlab{}.
\newblock \showarticletitle{BDTEX: A GQM-based Bayesian approach for the
  detection of antipatterns}.
\newblock \bibinfo{journal}{\emph{Journal of Systems and Software}}
  \bibinfo{volume}{84}, \bibinfo{number}{4} (\bibinfo{year}{2011}),
  \bibinfo{pages}{559--572}.
\newblock


\bibitem[\protect\citeauthoryear{{Khumnin} and {Senivongse}}{{Khumnin} and
  {Senivongse}}{2017}]%
        {Khumnin2017}
\bibfield{author}{\bibinfo{person}{P. {Khumnin}} {and} \bibinfo{person}{T.
  {Senivongse}}.} \bibinfo{year}{2017}\natexlab{}.
\newblock \showarticletitle{SQL antipatterns detection and database refactoring
  process}. In \bibinfo{booktitle}{\emph{2017 18th IEEE/ACIS International
  Conference on Software Engineering, Artificial Intelligence, Networking and
  Parallel/Distributed Computing (SNPD)}}. \bibinfo{pages}{199--205}.
\newblock
\showISSN{null}


\bibitem[\protect\citeauthoryear{Kim, Whitehead~Jr, and Zhang}{Kim
  et~al\mbox{.}}{2008}]%
        {kim2008classifying}
\bibfield{author}{\bibinfo{person}{Sunghun Kim}, \bibinfo{person}{E~James
  Whitehead~Jr}, {and} \bibinfo{person}{Yi Zhang}.}
  \bibinfo{year}{2008}\natexlab{}.
\newblock \showarticletitle{Classifying software changes: Clean or buggy?}
\newblock \bibinfo{journal}{\emph{IEEE Transactions on Software Engineering}}
  \bibinfo{volume}{34}, \bibinfo{number}{2} (\bibinfo{year}{2008}),
  \bibinfo{pages}{181--196}.
\newblock


\bibitem[\protect\citeauthoryear{Li and Shatnawi}{Li and Shatnawi}{2007}]%
        {li2007empirical}
\bibfield{author}{\bibinfo{person}{Wei Li} {and} \bibinfo{person}{Raed
  Shatnawi}.} \bibinfo{year}{2007}\natexlab{}.
\newblock \showarticletitle{An empirical study of the bad smells and class
  error probability in the post-release object-oriented system evolution}.
\newblock \bibinfo{journal}{\emph{Journal of Systems and Software}}
  \bibinfo{volume}{80}, \bibinfo{number}{7} (\bibinfo{year}{2007}),
  \bibinfo{pages}{1120--1128}.
\newblock


\bibitem[\protect\citeauthoryear{Ltd.}{Ltd.}{2014}]%
        {RedGateSQLSmells}
\bibfield{author}{\bibinfo{person}{Red Gate~Software Ltd.}}
  \bibinfo{year}{2014}\natexlab{}.
\newblock \bibinfo{title}{119 SQL Code Smells}.
\newblock
\newblock


\bibitem[\protect\citeauthoryear{{Lyu}, {Alotaibi}, and {Halfond}}{{Lyu}
  et~al\mbox{.}}{2019}]%
        {Lyu2019}
\bibfield{author}{\bibinfo{person}{Y. {Lyu}}, \bibinfo{person}{A. {Alotaibi}},
  {and} \bibinfo{person}{W.~G.~J. {Halfond}}.} \bibinfo{year}{2019}\natexlab{}.
\newblock \showarticletitle{Quantifying the Performance Impact of {SQL}
  Antipatterns on Mobile Applications}. In \bibinfo{booktitle}{\emph{2019 IEEE
  International Conference on Software Maintenance and Evolution (ICSME)}}.
  \bibinfo{pages}{53--64}.
\newblock
\showISSN{1063-6773}


\bibitem[\protect\citeauthoryear{{Macia}, {Arcoverde}, {Garcia}, {Chavez}, and
  {von Staa}}{{Macia} et~al\mbox{.}}{2012}]%
        {Macia2012}
\bibfield{author}{\bibinfo{person}{I. {Macia}}, \bibinfo{person}{R.
  {Arcoverde}}, \bibinfo{person}{A. {Garcia}}, \bibinfo{person}{C. {Chavez}},
  {and} \bibinfo{person}{A. {von Staa}}.} \bibinfo{year}{2012}\natexlab{}.
\newblock \showarticletitle{On the Relevance of Code Anomalies for Identifying
  Architecture Degradation Symptoms}. In \bibinfo{booktitle}{\emph{2012 16th
  European Conference on Software Maintenance and Reengineering}}.
  \bibinfo{pages}{277--286}.
\newblock
\showISSN{1534-5351}


\bibitem[\protect\citeauthoryear{M{\"a}ntyl{\"a} and Lassenius}{M{\"a}ntyl{\"a}
  and Lassenius}{2006}]%
        {Mantyla2006}
\bibfield{author}{\bibinfo{person}{Mika~V. M{\"a}ntyl{\"a}} {and}
  \bibinfo{person}{Casper Lassenius}.} \bibinfo{year}{2006}\natexlab{}.
\newblock \showarticletitle{Subjective evaluation of software evolvability
  using code smells: An empirical study}.
\newblock \bibinfo{journal}{\emph{Empirical Software Engineering}}
  \bibinfo{volume}{11}, \bibinfo{number}{3} (\bibinfo{date}{01 Sep}
  \bibinfo{year}{2006}), \bibinfo{pages}{395--431}.
\newblock
\showISSN{1573-7616}


\bibitem[\protect\citeauthoryear{Meurice, Nagy, and Cleve}{Meurice
  et~al\mbox{.}}{2016}]%
        {meurice2016static}
\bibfield{author}{\bibinfo{person}{Loup Meurice}, \bibinfo{person}{Csaba Nagy},
  {and} \bibinfo{person}{Anthony Cleve}.} \bibinfo{year}{2016}\natexlab{}.
\newblock \showarticletitle{Static analysis of dynamic database usage in {J}ava
  systems}. In \bibinfo{booktitle}{\emph{International Conference on Advanced
  Information Systems Engineering}}. Springer, \bibinfo{pages}{491--506}.
\newblock


\bibitem[\protect\citeauthoryear{Miller~Jr}{Miller~Jr}{2011}]%
        {miller2011survival}
\bibfield{author}{\bibinfo{person}{Rupert~G Miller~Jr}.}
  \bibinfo{year}{2011}\natexlab{}.
\newblock \bibinfo{booktitle}{\emph{Survival analysis}}.
  Vol.~\bibinfo{volume}{66}.
\newblock \bibinfo{publisher}{John Wiley \& Sons}.
\newblock


\bibitem[\protect\citeauthoryear{Mockus and Votta}{Mockus and Votta}{2000}]%
        {mockus2000identifying}
\bibfield{author}{\bibinfo{person}{Audris Mockus} {and}
  \bibinfo{person}{Lawrence~G Votta}.} \bibinfo{year}{2000}\natexlab{}.
\newblock \showarticletitle{Identifying Reasons for Software Changes using
  Historic Databases.}. In \bibinfo{booktitle}{\emph{Proc. of the 2000
  International Conference on Software Maintenance}}.
  \bibinfo{pages}{120--130}.
\newblock


\bibitem[\protect\citeauthoryear{Moha, Gu{\'e}h{\'e}neuc, Le~Meur, Duchien, and
  Tiberghien}{Moha et~al\mbox{.}}{2010}]%
        {moha2010domain}
\bibfield{author}{\bibinfo{person}{Naouel Moha}, \bibinfo{person}{Yann-Ga{\"e}l
  Gu{\'e}h{\'e}neuc}, \bibinfo{person}{Anne-Fran{\c{c}}oise Le~Meur},
  \bibinfo{person}{Laurence Duchien}, {and} \bibinfo{person}{Alban
  Tiberghien}.} \bibinfo{year}{2010}\natexlab{}.
\newblock \showarticletitle{From a domain analysis to the specification and
  detection of code and design smells}.
\newblock \bibinfo{journal}{\emph{Formal Aspects of Computing}}
  \bibinfo{volume}{22}, \bibinfo{number}{3-4} (\bibinfo{year}{2010}),
  \bibinfo{pages}{345--361}.
\newblock


\bibitem[\protect\citeauthoryear{Muse}{Muse}{2020}]%
        {replication}
\bibfield{author}{\bibinfo{person}{Biruk~Asmare Muse}.}
  \bibinfo{year}{2020}\natexlab{}.
\newblock \bibinfo{title}{Replication package}.
\newblock
\newblock
\urldef\tempurl%
\url{https://github.com/Biruk-Asmare/MSR_2020_SQLSmells_Prevalence}
\showURL{%
\tempurl}


\bibitem[\protect\citeauthoryear{Nagy and Cleve}{Nagy and Cleve}{2017}]%
        {nagy2017static}
\bibfield{author}{\bibinfo{person}{Csaba Nagy} {and} \bibinfo{person}{Anthony
  Cleve}.} \bibinfo{year}{2017}\natexlab{}.
\newblock \showarticletitle{A static code smell detector for {SQL} queries
  embedded in {J}ava code}. In \bibinfo{booktitle}{\emph{2017 IEEE 17th
  International Working Conference on Source Code Analysis and Manipulation
  (SCAM)}}. IEEE, \bibinfo{pages}{147--152}.
\newblock


\bibitem[\protect\citeauthoryear{Nagy and Cleve}{Nagy and Cleve}{2018}]%
        {nagy2018sqlinspect}
\bibfield{author}{\bibinfo{person}{Csaba Nagy} {and} \bibinfo{person}{Anthony
  Cleve}.} \bibinfo{year}{2018}\natexlab{}.
\newblock \showarticletitle{{SQLInspect}: A static analyzer to inspect database
  usage in {J}ava applications}. In \bibinfo{booktitle}{\emph{Proceedings of
  the 40th International Conference on Software Engineering: Companion
  Proceeedings}}. ACM, \bibinfo{pages}{93--96}.
\newblock


\bibitem[\protect\citeauthoryear{Olbrich, Cruzes, Basili, and Zazworka}{Olbrich
  et~al\mbox{.}}{2009}]%
        {olbrich2009evolution}
\bibfield{author}{\bibinfo{person}{Steffen Olbrich}, \bibinfo{person}{Daniela~S
  Cruzes}, \bibinfo{person}{Victor Basili}, {and} \bibinfo{person}{Nico
  Zazworka}.} \bibinfo{year}{2009}\natexlab{}.
\newblock \showarticletitle{The evolution and impact of code smells: A case
  study of two open source systems}. In \bibinfo{booktitle}{\emph{Proc. of the
  2009 3rd International Symposium on Empirical Software Engineering and
  Measurement}}. IEEE, \bibinfo{pages}{390--400}.
\newblock


\bibitem[\protect\citeauthoryear{{Olbrich}, {Cruzes}, and {Sjøberg}}{{Olbrich}
  et~al\mbox{.}}{2010}]%
        {Olbrich2010}
\bibfield{author}{\bibinfo{person}{S.~M. {Olbrich}}, \bibinfo{person}{D.~S.
  {Cruzes}}, {and} \bibinfo{person}{D.~I.~K. {Sjøberg}}.}
  \bibinfo{year}{2010}\natexlab{}.
\newblock \showarticletitle{Are all code smells harmful? A study of God Classes
  and Brain Classes in the evolution of three open source systems}. In
  \bibinfo{booktitle}{\emph{Proc. of the 2010 IEEE International Conference on
  Software Maintenance}}. \bibinfo{pages}{1--10}.
\newblock
\showISSN{1063-6773}


\bibitem[\protect\citeauthoryear{Palomba}{Palomba}{2015}]%
        {palomba2015textual}
\bibfield{author}{\bibinfo{person}{Fabio Palomba}.}
  \bibinfo{year}{2015}\natexlab{}.
\newblock \showarticletitle{Textual analysis for code smell detection}. In
  \bibinfo{booktitle}{\emph{Proceedings of the 37th International Conference on
  Software Engineering-Volume 2}}. IEEE Press, \bibinfo{pages}{769--771}.
\newblock


\bibitem[\protect\citeauthoryear{Palomba, Bavota, Di~Penta, Fasano, Oliveto,
  and De~Lucia}{Palomba et~al\mbox{.}}{2018}]%
        {palomba2018diffuseness}
\bibfield{author}{\bibinfo{person}{Fabio Palomba}, \bibinfo{person}{Gabriele
  Bavota}, \bibinfo{person}{Massimiliano Di~Penta}, \bibinfo{person}{Fausto
  Fasano}, \bibinfo{person}{Rocco Oliveto}, {and} \bibinfo{person}{Andrea
  De~Lucia}.} \bibinfo{year}{2018}\natexlab{}.
\newblock \showarticletitle{On the diffuseness and the impact on
  maintainability of code smells: a large scale empirical investigation}.
\newblock \bibinfo{journal}{\emph{Empirical Software Engineering}}
  \bibinfo{volume}{23}, \bibinfo{number}{3} (\bibinfo{year}{2018}),
  \bibinfo{pages}{1188--1221}.
\newblock


\bibitem[\protect\citeauthoryear{{Palomba}, {Bavota}, {Penta}, {Oliveto}, and
  {Lucia}}{{Palomba} et~al\mbox{.}}{2014}]%
        {Palomba2014}
\bibfield{author}{\bibinfo{person}{F. {Palomba}}, \bibinfo{person}{G.
  {Bavota}}, \bibinfo{person}{M.~D. {Penta}}, \bibinfo{person}{R. {Oliveto}},
  {and} \bibinfo{person}{A.~D. {Lucia}}.} \bibinfo{year}{2014}\natexlab{}.
\newblock \showarticletitle{Do They Really Smell Bad? A Study on Developers'
  Perception of Bad Code Smells}. In \bibinfo{booktitle}{\emph{2014 IEEE
  International Conference on Software Maintenance and Evolution}}.
  \bibinfo{pages}{101--110}.
\newblock
\showISSN{1063-6773}


\bibitem[\protect\citeauthoryear{Peters and Zaidman}{Peters and
  Zaidman}{2012}]%
        {peters2012evaluating}
\bibfield{author}{\bibinfo{person}{Ralph Peters} {and} \bibinfo{person}{Andy
  Zaidman}.} \bibinfo{year}{2012}\natexlab{}.
\newblock \showarticletitle{Evaluating the lifespan of code smells using
  software repository mining}. In \bibinfo{booktitle}{\emph{2012 16th European
  Conference on Software Maintenance and Reengineering}}. IEEE,
  \bibinfo{pages}{411--416}.
\newblock


\bibitem[\protect\citeauthoryear{Peto and Peto}{Peto and Peto}{1972}]%
        {peto1972asymptotically}
\bibfield{author}{\bibinfo{person}{Richard Peto} {and} \bibinfo{person}{Julian
  Peto}.} \bibinfo{year}{1972}\natexlab{}.
\newblock \showarticletitle{Asymptotically efficient rank invariant test
  procedures}.
\newblock \bibinfo{journal}{\emph{Journal of the Royal Statistical Society:
  Series A (General)}} \bibinfo{volume}{135}, \bibinfo{number}{2}
  (\bibinfo{year}{1972}), \bibinfo{pages}{185--198}.
\newblock


\bibitem[\protect\citeauthoryear{Piatetsky-Shapiro}{Piatetsky-Shapiro}{1991}]%
        {piatetsky1991discovery}
\bibfield{author}{\bibinfo{person}{Gregory Piatetsky-Shapiro}.}
  \bibinfo{year}{1991}\natexlab{}.
\newblock \showarticletitle{Discovery, analysis, and presentation of strong
  rules}.
\newblock \bibinfo{journal}{\emph{Knowledge discovery in databases}}
  (\bibinfo{year}{1991}), \bibinfo{pages}{229--238}.
\newblock


\bibitem[\protect\citeauthoryear{Riel}{Riel}{1996}]%
        {Riel1996}
\bibfield{author}{\bibinfo{person}{Arthur~J. Riel}.}
  \bibinfo{year}{1996}\natexlab{}.
\newblock \bibinfo{booktitle}{\emph{Object-Oriented Design Heuristics}
  (\bibinfo{edition}{1st} ed.)}.
\newblock \bibinfo{publisher}{Addison-Wesley Longman Publishing Co., Inc.},
  \bibinfo{address}{USA}.
\newblock
\showISBNx{020163385X}


\bibitem[\protect\citeauthoryear{Rodr{\'\i}guez-P{\'e}rez, Robles, and
  Gonz{\'a}lez-Barahona}{Rodr{\'\i}guez-P{\'e}rez et~al\mbox{.}}{2018}]%
        {rodriguez2018reproducibility}
\bibfield{author}{\bibinfo{person}{Gema Rodr{\'\i}guez-P{\'e}rez},
  \bibinfo{person}{Gregorio Robles}, {and} \bibinfo{person}{Jes{\'u}s~M
  Gonz{\'a}lez-Barahona}.} \bibinfo{year}{2018}\natexlab{}.
\newblock \showarticletitle{Reproducibility and credibility in empirical
  software engineering: A case study based on a systematic literature review of
  the use of the szz algorithm}.
\newblock \bibinfo{journal}{\emph{Information and Software Technology}}
  \bibinfo{volume}{99} (\bibinfo{year}{2018}), \bibinfo{pages}{164--176}.
\newblock


\bibitem[\protect\citeauthoryear{Santos, Rocha-Junior, Prates, do~Nascimento,
  Freitas, and de~Mendonça}{Santos et~al\mbox{.}}{2018}]%
        {Santos2018}
\bibfield{author}{\bibinfo{person}{José Amancio~M. Santos},
  \bibinfo{person}{João~B. Rocha-Junior}, \bibinfo{person}{Luciana Carla~Lins
  Prates}, \bibinfo{person}{Rogeres~Santos do Nascimento},
  \bibinfo{person}{Mydiã~Falcão Freitas}, {and} \bibinfo{person}{Manoel~Gomes
  de Mendonça}.} \bibinfo{year}{2018}\natexlab{}.
\newblock \showarticletitle{A systematic review on the code smell effect}.
\newblock \bibinfo{journal}{\emph{Journal of Systems and Software}}
  \bibinfo{volume}{144} (\bibinfo{year}{2018}), \bibinfo{pages}{450 -- 477}.
\newblock
\showISSN{0164-1212}


\bibitem[\protect\citeauthoryear{{Sharma}, {Fragkoulis}, {Rizou}, {Bruntink},
  and {Spinellis}}{{Sharma} et~al\mbox{.}}{2018}]%
        {Sharma2018}
\bibfield{author}{\bibinfo{person}{T. {Sharma}}, \bibinfo{person}{M.
  {Fragkoulis}}, \bibinfo{person}{S. {Rizou}}, \bibinfo{person}{M. {Bruntink}},
  {and} \bibinfo{person}{D. {Spinellis}}.} \bibinfo{year}{2018}\natexlab{}.
\newblock \showarticletitle{Smelly Relations: Measuring and Understanding
  Database Schema Quality}. In \bibinfo{booktitle}{\emph{2018 IEEE/ACM 40th
  International Conference on Software Engineering: Software Engineering in
  Practice Track (ICSE-SEIP)}}. \bibinfo{pages}{55--64}.
\newblock
\showISSN{null}


\bibitem[\protect\citeauthoryear{Shatnawi and Li}{Shatnawi and Li}{2006}]%
        {shatnawi2006investigation}
\bibfield{author}{\bibinfo{person}{Raed Shatnawi} {and} \bibinfo{person}{Wei
  Li}.} \bibinfo{year}{2006}\natexlab{}.
\newblock \showarticletitle{An investigation of bad smells in object-oriented
  design}. In \bibinfo{booktitle}{\emph{Third International Conference on
  Information Technology: New Generations (ITNG'06)}}. IEEE,
  \bibinfo{pages}{161--165}.
\newblock


\bibitem[\protect\citeauthoryear{{Sjøberg}, {Yamashita}, {Anda}, {Mockus}, and
  {Dybå}}{{Sjøberg} et~al\mbox{.}}{2013}]%
        {Sjoberg2013}
\bibfield{author}{\bibinfo{person}{D.~I.~K. {Sjøberg}}, \bibinfo{person}{A.
  {Yamashita}}, \bibinfo{person}{B.~C.~D. {Anda}}, \bibinfo{person}{A.
  {Mockus}}, {and} \bibinfo{person}{T. {Dybå}}.}
  \bibinfo{year}{2013}\natexlab{}.
\newblock \showarticletitle{Quantifying the Effect of Code Smells on
  Maintenance Effort}.
\newblock \bibinfo{journal}{\emph{IEEE Transactions on Software Engineering}}
  \bibinfo{volume}{39}, \bibinfo{number}{8} (\bibinfo{date}{Aug}
  \bibinfo{year}{2013}), \bibinfo{pages}{1144--1156}.
\newblock
\showISSN{2326-3881}


\bibitem[\protect\citeauthoryear{{\'S}liwerski, Zimmermann, and
  Zeller}{{\'S}liwerski et~al\mbox{.}}{2005}]%
        {sliwerski2005changes}
\bibfield{author}{\bibinfo{person}{Jacek {\'S}liwerski},
  \bibinfo{person}{Thomas Zimmermann}, {and} \bibinfo{person}{Andreas Zeller}.}
  \bibinfo{year}{2005}\natexlab{}.
\newblock \showarticletitle{When do changes induce fixes?}. In
  \bibinfo{booktitle}{\emph{ACM sigsoft software engineering notes}},
  Vol.~\bibinfo{volume}{30}. ACM, \bibinfo{pages}{1--5}.
\newblock


\bibitem[\protect\citeauthoryear{Spadini, Aniche, and Bacchelli}{Spadini
  et~al\mbox{.}}{2018}]%
        {PyDriller}
\bibfield{author}{\bibinfo{person}{Davide Spadini}, \bibinfo{person}{Maurício
  Aniche}, {and} \bibinfo{person}{Alberto Bacchelli}.}
  \bibinfo{year}{2018}\natexlab{}.
\newblock \showarticletitle{{PyDriller}: Python Framework for Mining Software
  Repositories}. In \bibinfo{booktitle}{\emph{Proc of the 26th ACM Joint
  European Software Engineering Conference and Symposium on the Foundations of
  Software Engineering (ESEC/FSE)}}. \bibinfo{pages}{908–911}.
\newblock


\bibitem[\protect\citeauthoryear{Tsantalis}{Tsantalis}{2010}]%
        {tsantalis2010evaluation}
\bibfield{author}{\bibinfo{person}{Nikolaos Tsantalis}.}
  \bibinfo{year}{2010}\natexlab{}.
\newblock \showarticletitle{Evaluation and improvement of software
  architecture: Identification of design problems in object-oriented systems
  and resolution through refactorings}.
\newblock \bibinfo{journal}{\emph{Diss. Ph. D. dissertation, Univ. of
  Macedonia}} (\bibinfo{year}{2010}).
\newblock


\bibitem[\protect\citeauthoryear{Yamashita and Moonen}{Yamashita and
  Moonen}{2013}]%
        {Yamashita2013}
\bibfield{author}{\bibinfo{person}{Aiko Yamashita} {and} \bibinfo{person}{Leon
  Moonen}.} \bibinfo{year}{2013}\natexlab{}.
\newblock \showarticletitle{Exploring the Impact of Inter-Smell Relations on
  Software Maintainability: An Empirical Study}. In
  \bibinfo{booktitle}{\emph{Proceedings of the 2013 International Conference on
  Software Engineering}} \emph{(\bibinfo{series}{ICSE ’13})}.
  \bibinfo{publisher}{IEEE Press}, \bibinfo{pages}{682–691}.
\newblock
\showISBNx{9781467330763}


\bibitem[\protect\citeauthoryear{Zazworka, Shaw, Shull, and Seaman}{Zazworka
  et~al\mbox{.}}{2011}]%
        {zazworka2011investigating}
\bibfield{author}{\bibinfo{person}{Nico Zazworka}, \bibinfo{person}{Michele~A
  Shaw}, \bibinfo{person}{Forrest Shull}, {and} \bibinfo{person}{Carolyn
  Seaman}.} \bibinfo{year}{2011}\natexlab{}.
\newblock \showarticletitle{Investigating the impact of design debt on software
  quality}. In \bibinfo{booktitle}{\emph{Proceedings of the 2nd Workshop on
  Managing Technical Debt}}. ACM, \bibinfo{pages}{17--23}.
\newblock


\end{thebibliography}

\end{document}